\documentclass[12pt]{article}
\usepackage{amssymb}
\usepackage{amsmath}
\usepackage{latexsym}
\usepackage[usenames]{color}
\usepackage{fancybox}
\usepackage{simplewick}
\usepackage{cite}
\usepackage{framed}
\definecolor{shadecolor}{rgb}{0.9,0.9,0.95}
\usepackage{setspace}
\definecolor{refkey}{rgb}{0.5,0.5,0}
\definecolor{labelkey}{rgb}{0.5,0.5,0}
\definecolor{citekey}{rgb}{0.5,0.5,0}
\definecolor{darkgreen}{rgb}{0,0.5,0}
\definecolor{darkblue}{cmyk}{0.9,0.9,0,0}
\definecolor{darkred}{rgb}{0.6,0,0.3}

\usepackage{ifpdf}
\ifpdf
\usepackage{graphicx} 
\usepackage{epsfig}
\usepackage[setpagesize=false,pagebackref=false, linktocpage, bookmarksopen=true, colorlinks=true, linkcolor=darkblue,citecolor=darkblue,urlcolor=darkblue]{hyperref}
\usepackage{hyperref}
\newcommand{\arXiv}[2]{\href{http://arxiv.org/abs/#1}{{\tt arXiv:#2}}}
\newcommand{\hep}[2]{\href{http://arxiv.org/abs/#1}{{\tt #2}}}

\else
\usepackage[dvipdfmx]{graphicx}
\newcommand{\arXiv}[2]{[{\tt arXiv:#2}]}
\newcommand{\hep}[2]{[{\tt #2}]}

\fi

\renewcommand{\thefootnote}{\arabic{footnote}}

\def\eqref#1{(\ref{#1})}

\def\beq{\begin{equation}}
\def\eeq{\end{equation}}

\begin{document}
\thispagestyle{empty}

\renewcommand{\thefootnote}{\fnsymbol{footnote}}
\setcounter{page}{1}
\setcounter{footnote}{0}
\setcounter{figure}{0}
\begin{flushright}
UT-Komaba 17-3
\end{flushright}
\begin{center}
$$$$
{\Large\textbf{\mathversion{bold}
Structure constants of operators on the Wilson loop from integrability
}\par}

\vspace{1.2cm}

\textrm{Minkyoo Kim$^{1,2}$ \,\ and \,\ Naoki Kiryu $^{3}$}
\\ \vspace{1.2cm}
\footnotesize{\textit{
$^{1}$National Institute for Theoretical Physics, School of Physics and Mandelstam Institute for Theoretical Physics, University of the Witwatersrand, Johannesburg
Wits 2050, South Africa\\
$^{2}$MTA Lendulet Holographic QFT Group, Wigner Research Centre, Budapest 114, P.O.B. 49, Hungary\\
$^{3}$Institute of Physics, University of Tokyo, Komaba, Meguro-ku, Tokyo 153-8902 Japan\\
}  
\vspace{6mm}
}
\textrm{E-mail: minkyoo.kim@wits.ac.za , kiryu@hep1.c.u-tokyo.ac.jp}

\par\vspace{1.5cm}

\textbf{Abstract}\vspace{2mm}
\end{center}
We study structure constants of local operators inserted on the Wilson loop in ${\cal N}=4$ super Yang-Mills theory. 
We conjecture the finite coupling expression of the structure constant which is interpreted as one hexagon with three mirror edges contracted by the boundary states. 
This is consistent with a holographic description of the correlator as the cubic open string vertex which consists of one hexagonal patch and three boundaries. We check its validity at the weak coupling where the asymptotic expression reduces to the summation over all possible ways of changing the signs of magnon momenta in the hexagon form factor. 
For this purpose, we compute the structure constants in the $SU(2)$ sector at tree level using the correspondence between operators on the Wilson loop and the open spin chain. The result is nicely matched with our conjecture at the weak coupling regime.
\noindent

\setcounter{page}{1}
\renewcommand{\thefootnote}{\arabic{footnote}}
\setcounter{footnote}{0}
\setcounter{tocdepth}{2}
\newpage
\tableofcontents

\parskip 5pt plus 1pt   \jot = 1.5ex

\newpage
\section{Introduction}

The $AdS/CFT$ correspondence is one of the most interesting ideas which has appeared in string theory \cite{AdS/CFT}. 
The prototypical example of the correspondence is the duality between type IIB superstring theory on $AdS_{5} \times S^{5}$ and ${\cal N}=4$ supersymmetric Yang-Mills(${\cal N}=4$ SYM) theory in ${\mathbb R}^{(1,3)}$ which is defined on the boundary of $AdS$. 

The ${\cal N}=4$ SYM theory has the maximal number of supersymmetries in four dimensional spacetime. Furthermore, since the theory is a conformal field theory, the main dynamical building blocks of correlation functions are conformal dimensions and structure constants of gauge invariant operators. 
On the other hand, such observables correspond to energies and interaction vertices of strings propagating on $AdS$ spacetime.
However, since the duality has a strong/weak property, we should be able to compute nonperturbatively on either side to verify it. 
This is a nice property to use, but also a cause of difficulty. 
Because of this reason, the earliest works have focused on BPS operators since there operators do not have quantum corrections.  

However, the crucial idea which led to numerous developments in understanding the $AdS/CFT$ correspondence beyond the protected operators was the integrable structure found on both sides of the planar $AdS/CFT$ \cite{Beisert:2010jr}.\footnote{Recently, a non-integrable but solvable model of holography was proposed \cite{syk1, syk2}. The SYK model seems to provide an another interesting example to understand the $AdS/CFT$ duality.} 
Using integrability, one could analytically determine physical observables such as the exact $S$-matrix \cite{Beisert:2005tm, Arutyunov:2006yd},  asymptotic Bethe ansatz equations \cite{Beisert:2005fw} and finite size energy shifts of local operators or solitonic strings \cite{Arutyunov:2006gs, Janik:2007wt, Bajnok:2008bm}.   
In addition, a resummation formula which nonperturbatively includes all finite size effects was finally proposed in a set of finite nonlinear integral equations called the quantum spectral curves \cite{Gromov:2014caa}. With this formulation one can in principle calculate the anomalous dimensions of all gauge invariant operators or the energies of quantum strings with any desired accuracy.\footnote{For example, see \cite{Marboe:2014sya} for understanding how the quantum spectral curve works.} For determining spectra, such exact methods were surprisingly matched with all perturbation results till now.

On the other hand, the three-point function problem of gauge invariant operators is not completely formulated yet. 
Nevertheless, there were some seminal works which suggested breakthroughs to understand the three point function.
On the gauge theory side, the three-point functions of gauge invariant operators were calculated systematically using integrability techniques as, the so called tailoring method in \cite{Escobedo:2010xs}.\footnote{The pioneer papers in the computation of the three-point function are \cite{Okuyama:2004bd, Roiban:2004va, Alday:2005nd}. The tailoring method is based on these previous papers.}
On the string theory side, the holographic three-point functions were also unveiled in \cite{Zarembo:2010rr, Costa:2010rz, Janik:2011bd, Kazama:2011cp, Kazama:2012is
, Kazama:2013qsa}.\footnote{Some related works for the three-point function are done in \cite{Kazama:2014sxa, Jiang:2014cya, Kazama:2015iua, Jiang:2015lda, Kazama:2016cfl, Jiang:2016ulr}.}

A great development has recently been suggested in computing three-point correlators of closed strings or single trace operators at finite coupling \cite{BKV}.
The key idea of \cite{BKV} is to cut the string pants diagram into the two hexagon form factors which could be thought as new fundamental building blocks in correlation functions.\footnote{The different approach for computing correlators based on integrable bootstrap was developed in \cite{Bajnok:2015hla, Bajnok:2015ftj, Bajnok:2017mdf}. In the development, as the worldsheet diagram is differently cut, the main building block is considered by an octagon form factor or a decompactified string field theory vertex.} Beyond asymptotic level,\footnote{Here, {\it asymptotic} means that all spin-chain lengths $L_{i}$ and all bridge lengths $\ell_{ij}=(L_{i}+L_{j}-L_{k})/2$ are taken to be infinite. Then finite size corrections are ignored.} the hexagon idea was recently generalized to a few wrapping orders \cite{Eden:2015ija, Basso:2015eqa, Basso:2017muf} and to higher-point functions \cite{Fleury:2016ykk}.\footnote{The asymptotic four-point function based on the usual operator product expansion also appeared in \cite{Basso:2017khq}}.

According to the hexagon decomposition method, the structure constants are constructed by two procedures such as the asymptotic part (where the bridge lengths are infinite) and mirror particle corrections (which correspond to finite-size corrections):\footnote{The proposal is actually to treat infinitely many mirror particle corrections. Thus, we would still need a resummation formula which has exact finite size corrections to completely understand correlation functions as in spectral problem.}
\begin{equation}
C_{123} \sim \int_{\rm mirror} \sum_{\rm magnon} {\cal H} \times {\cal H},
\end{equation}
where ${\cal H}$ is called the hexagon form factor.

As a natural question, we would like to consider the open string version of the hexagon method.
Open strings are generally attached to $D$-branes, and their dual gauge invariant operators are known for various $D$-brane configurations.
In particular, some open string configurations have been shown to be integrable with specific boundary conditions \cite{Mann:2006rh, Correa:2008av, Correa:2011nz, Dekel:2011ja}.\footnote{For example, open strings attached to the maximal giant graviton are quantum integrable through exact reflection matrices which obeys the boundary Yang-Baxter equation \cite{Berenstein:2005vf, Hofman:2007xp}.} 

\begin{figure}[t]
\begin{center}
\includegraphics[width=6cm]{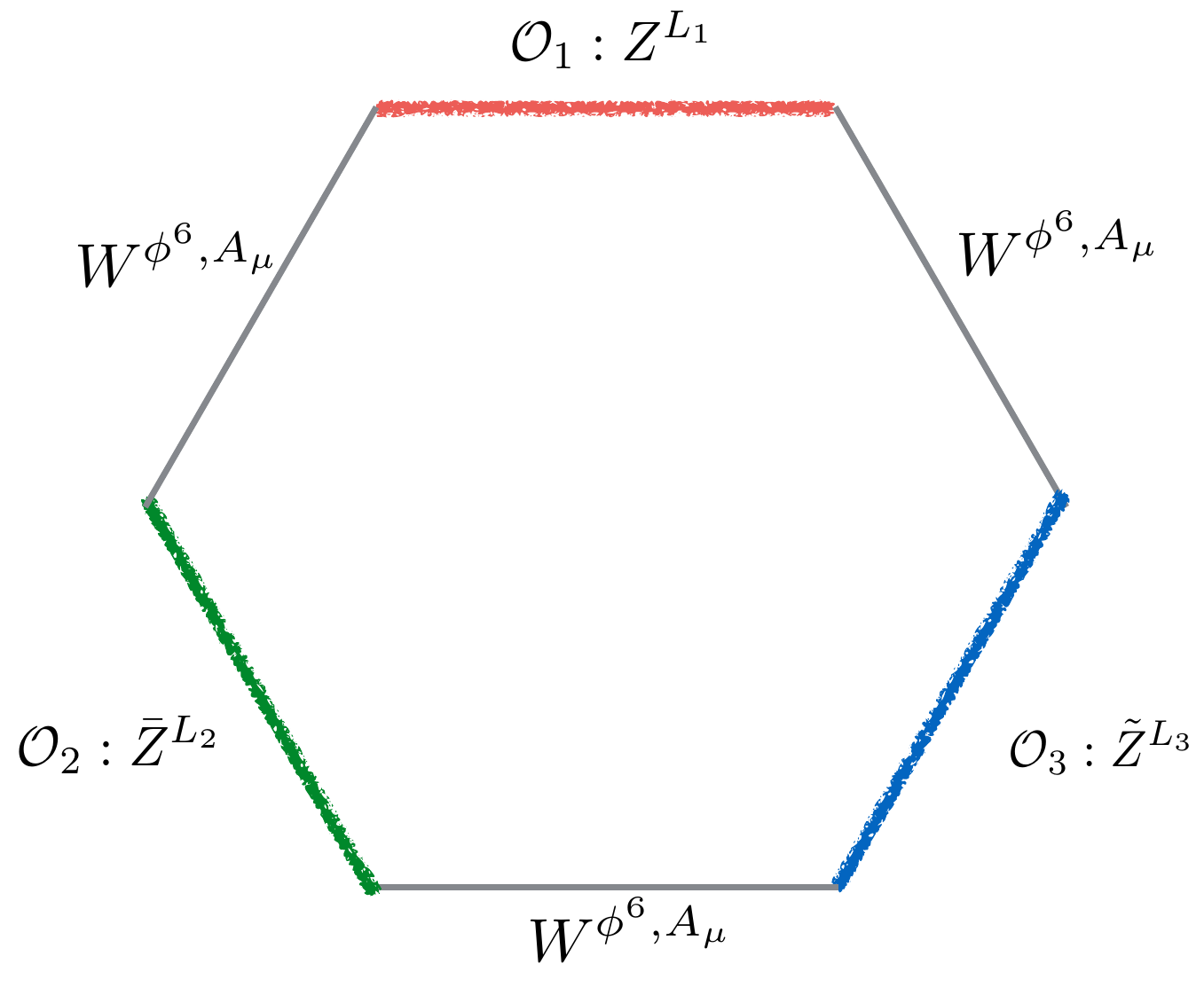}
\caption{Wilson loop with operator insertions. For a BPS configuration, we consider the following operators constructed by the complex scalars $Z=\phi_{1}+i\phi_{2}$ and $Y=\phi_{3}+i\phi_{4}$: ${\cal O}_{1}:Z^{L_{1}},\ \ {\cal O}_{2}:\bar{Z}^{L_{1}}$ and ${\cal O}_{3}:\tilde{Z}^{L_{3}}=(Z+\bar{Z}+Y-\bar{Y})^{L_{3}}$. Furthermore we choose $\phi_{6}$ as the coupled scalar into the Wilson loops, since there is no direct contraction between the inserted operators and the Wilson loop.}
\label{fig:config}
\end{center}
\end{figure}

In this paper, we shall focus on the three-point functions of open strings stretched to the $AdS$ boundary whose spectral problem was already studied using integrability techniques \cite{Drukker:2012de, Correa:2012hh}. 
Such an open string configuration is realized by three local operators inserted on the $1/2$-BPS Wilson loop in dual gauge theory.\footnote{Actually, we can construct a nontrivial Wilson loops configuration without local operator insertions. This is realized by ``defect changing operator'', which can change the scalar coupled to the Wilson loop. The Wilson loop configuration with the defect changing operators can be thought as open strings without physical magnons \cite{KKKN}. } Here the $1/2$-BPS Wilson loop is coupled to not only gauge fields but also to a scalar field. Owing to $SO(6)$ $R$-symmetry in ${\cal N}=4$ supersymmetry, the scalar field $\phi_{i}$ lives on a $6$-dimensional inner space (The $6$-dimensional unit vector is denoted by $n^{i}$). Now, we define a segment from $x_{i}$ to $x_{j}$ of the Wilson loops as
\begin{align}
W|_{x_{i}}^{x_{j}}=\exp\Big[\int_{x_{i}}^{x_{j}}d\tau(iA_{\mu}\dot{x}^{\mu}+\phi_{i}n^{i}|\dot{x}^{\mu}|)\Big],
\end{align}
which is not gauge invariant.
The three-point functions of the local operators (denoted by ${\cal O}_{i}$) inserted on the Wilson loops is now constructed by the three operators and three Wilson loop segments in figure \ref{fig:config}.
Namely, it is precisely the expectation value of the gauge invariant nonlocal operator :
\begin{align}
\langle W[{\cal O}_{1}(x_{1}){\cal O}_{2}(x_{2}){\cal O}_{3}(x_{3})]\rangle=\langle{\rm Tr}[W|_{x_{3}}^{x_{1}}\ {\cal O}_{1}(x_{1})\ W|_{x_{1}}^{x_{2}}\ {\cal O}_{2}(x_{2})\ W|_{x_{2}}^{x_{3}}\ {\cal O}_{3}(x_{3})]\rangle .
\end{align}
Since we expect the space-time dependence of the three-point functions to be determined by conformal symmetry,\footnote{If we align both the operators and segments of the Wilson loop with a straight line, the correlator is characterized by $SL(2,R)$ symmetry. Then the space-time dependence of such a correlator is completely fixed to an usual form of three-point functions of conformal field theory.}  
we have
\begin{align}
\langle W[{\cal O}_{1}(x_{1}){\cal O}_{2}(x_{2}){\cal O}_{3}(x_{3})]\rangle=\frac{C_{123}}{x_{12}^{\Delta_{1}+\Delta_{2}-\Delta_{3}}x_{23}^{\Delta_{2}+\Delta_{3}-\Delta_{1}}x_{31}^{\Delta_{3}+\Delta_{1}-\Delta_{2}}},
\end{align}
where $x_{ij}=x_{i}-x_{j}$. $\Delta_{i}$ and $C_{123}$ are the conformal dimension and the structure constant respectively\footnote{Several papers on correlation functions on the Wilson loop appeared recently in \cite{CDD,GRT}.}.

\begin{figure}[t]
\begin{center}
\includegraphics[width=10cm]{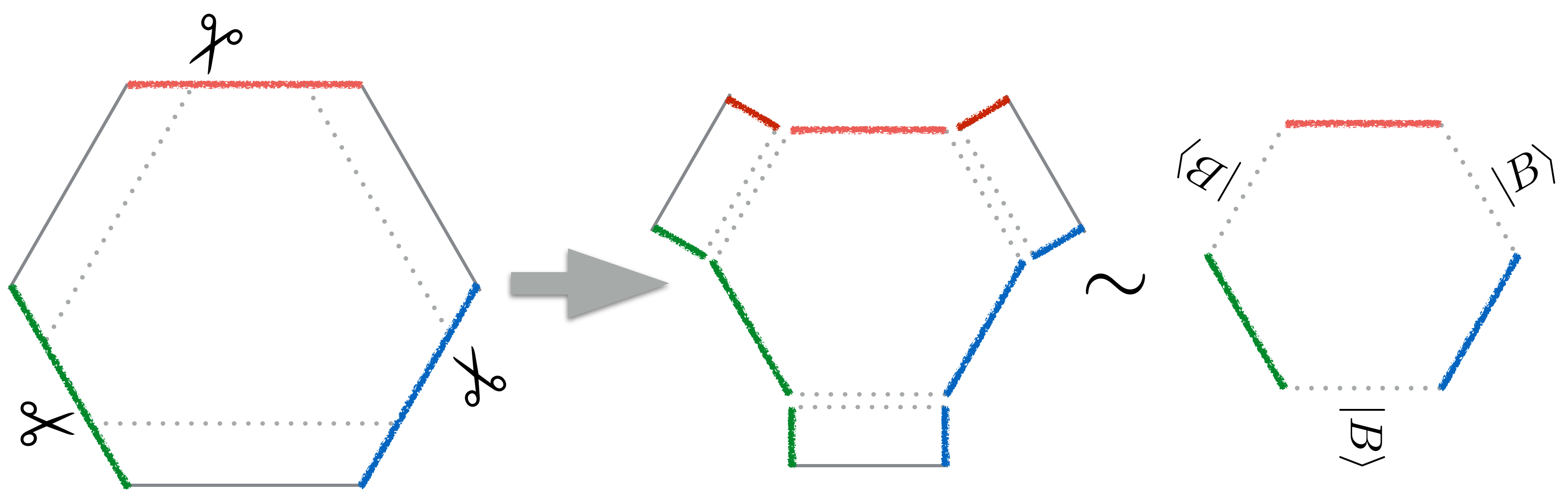}
\caption{Three-point function of the open strings (open spin-chains)}
\label{fig:oct}
\end{center}
\end{figure}

Pictorially, the three-point function of operators inserted on the Wilson loop is implemented by a hexagonal object. Cutting the three seams, the hexagonal object can be described by one hexagon with three mirror edges contracted by the boundary
states $|B\rangle$.\footnote{Up to the tree-level and the $SU(2)$ sector analysis, we could check that there is no additional object except the hexagon for bootstrap.} See figure \ref{fig:oct}.
The hexagon with three mirror edges can be taken as the well-known hexagon form factor \cite{BKV} by putting physical magnons into the hexagon twist operator. As our setup implies an open string version of the hexagon approach, we shall investigate how the open string three-point function is written by using the hexagon form factor and how it can reduce to an appropriate form in the dual weak coupling analysis.

The outline of this paper is as follows. 
In section \ref{sec2}, we suggest the finite coupling expression for structure constants of local operators on Wilson loops and explain how our suggestion is related to the hexagon form factor. 
To check validity of the conjecture, we perform the tree-level analysis of structure constants for the $SU(2)$ sector in section \ref{sec3}.
In addition, we explain that the weak coupling result is nicely matched with our conjecture.
We conclude with a discussion. A few appendices are provided for some technical details and helpful comments.

\section{Conjecture for structure constants at the finite coupling \label{sec2}}
As we explained in introduction, the asymptotically exact three point function of single trace operators which are dual to closed strings is provided by decomposing the closed string worldsheet diagram into two hexagon form factors with mirror particle dressing related to gluing two hexagon twist operators together \cite{BKV}.
Thus, we should consider all possible distributions of magnon momenta to two each hexagonal patches besides mirror particle summations for wrapping effects. The distributions are characterized through possible propagation factors and $S$-matrix factors.

On the other hand, for the three open strings worldsheet diagram, one would just have one hexagon with three mirror edges contracted by the boundary states which would simply be the ends of open strings related to $D$-branes. 
The resulting situation can be described by figure \ref{fig:oct} where the cutting just means that the boundary information should be glued to the hexagon.
There would be no bipartite partitions as in the closed string case since there is just a hexagon twist operator. 
However, we now have to consider reflection effects at boundaries where each open strings end, and we need to sum over all possible ways to put minus signs to magnons because magnons can reflect with minus sign at the boundaries. 

\begin{figure}[t]
\begin{center}
\includegraphics[width=11cm]{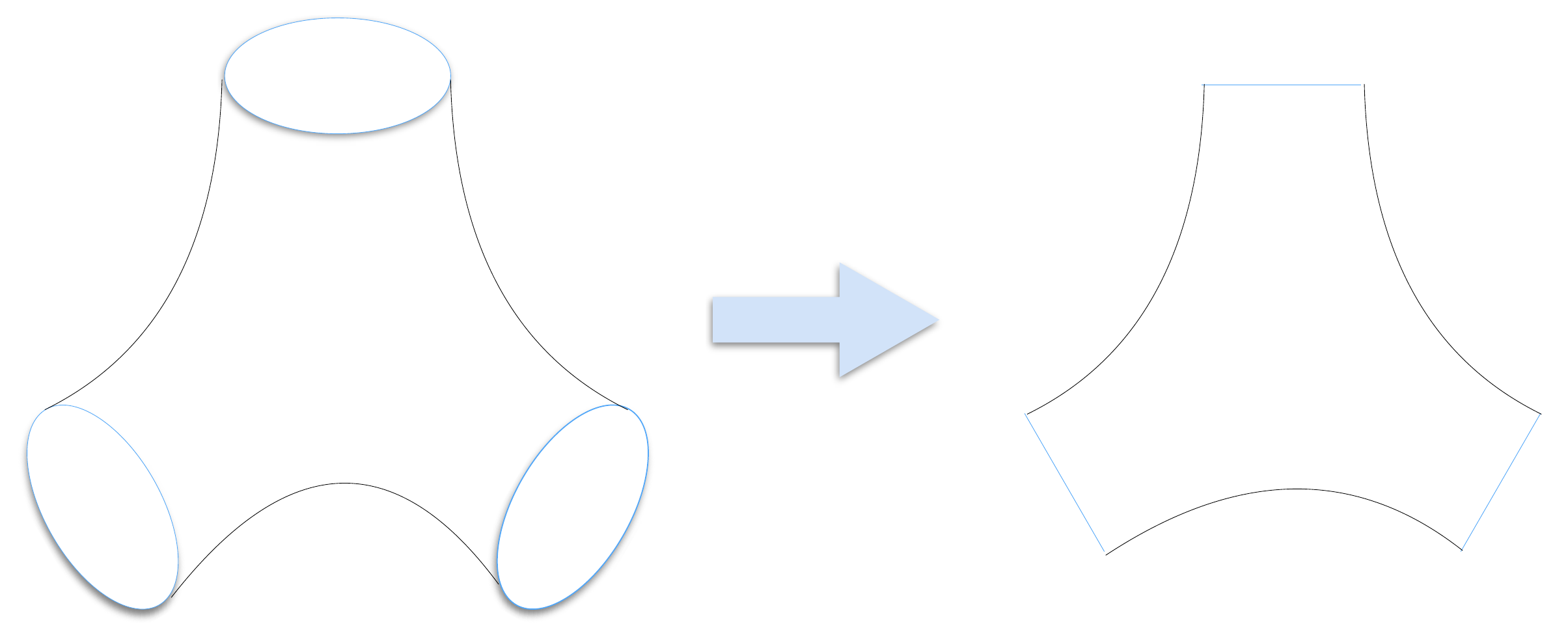}
\caption{Three closed strings worldsheet to three open strings worldsheet}
\label{fig:ows}
\end{center}
\end{figure}

Unlike in the three closed strings worldsheet as the pair of pants, the three open strings worldsheet diagram is given by a planar diagram with six edges as in figure \ref{fig:ows} where the three edges would correspond to ends of open strings.\footnote{Our setup does not have any $D$-brane since open strings stretch to $AdS$ boundary. However, as other integrable open strings end to some $D$-brane, these three edges would have information for such a $D$-brane. } 
The other three edges describe open strings themselves which are propagating in $AdS$ spacetime.

We then conjecture all-loop expression for the structure constant of operators on Wilson loops. 
For example, for finite coupling asymptotic structure constants of the $SU(2)$ sector with a nontrivial operator and two vacuum states, we suggest
\begin{align}
\left(\frac{C_{123}^{M\circ\circ}}{C_{123}^{\circ\circ\circ}}\right)^{2}=\frac{(e^{i(p_{1}+\cdots+p_{M})\ell_{12}}{\cal K}^{(M)})^{2}}{{\rm det}(\partial_{u_{i}}\phi_{j})\prod_{i<j} S(p_{j},p_{i})e^{i(p_{1}+\cdots+p_{M})L_{1}}}, \label{normmulti}
\end{align}
where the denominator is a norm part by a product of $S$-matrices times the open-chain version of Gaudin norm factor
which is a determinant of differential $\phi$ defined from\footnote{We explicitly wrote the left and the right reflection matrices in here. However, when we compute the tree-level structure constants, we shall take the half-step shifted basis. Then, the reflection factors just become $1$ which means the Neumann boundary conditions in the spin-chain.}
\begin{align}
e^{i\phi_{j}}\equiv e^{2ip_{j}L_{1}}\prod_{k\neq j}S(u_{k},u_{j}) B_{L}(p_{j})S(-u_{k},u_{j})B_{1}(-p_{j}) \label{byeq}
\end{align}
with respect to rapidity variables $u_{i}$ (not the momentum), and by the propagation factor for the length of the operator ${\cal O}_{1}$.
Also, we define 
\begin{align}
{\cal K}^{(M)}=\sum_{{\sf P}_{+}\cup {\sf P_-}=\{1,\ldots,M\}}\left[\prod_{k\in {\sf P}_{-}}\left(-e^{2i p_k \ell_{13}}B_{L}(p_{k}) \right)\prod_{l<k}S(p_k,p_l)S(-p_k,p_l)\right]\prod_{i<j}h_{YY}(\hat{p}_i,\hat{p}_j)
\end{align}
with
\begin{align}
\hat{p}_i=\begin{cases}p_i \qquad & i\in {\sf P}_{+}\\-p_i \qquad & i\in {\sf P}_{-}\end{cases}.
\end{align}
Note that $\ell_{ij}\equiv(L_{i}+L_{j}-L_{k})/2$ ($i,j$ and $k$ are all different ) is the bridge length between the operator ${\cal O}_{i}$ and ${\cal O}_{j}$. Thus, in the numerator there is a propagation factor for the bridge length between the operator ${\cal O}_{1}$ and ${\cal O}_{2}$.
The $B_{L}(p)$ is the reflection amplitude at finite coupling and $h_{YY}$ is the two-particle hexagon form factor at the finite coupling.\footnote{The multi-magnon hexagon form factor can be factorized into two-body hexagon form factor.}   
In the case of two magnon, the above expression is simply written as
\begin{align}
&{\cal K}^{(2)}=h_{YY}(u,v)-S(p_{2},p_{1})S(-p_{2},p_{1})B_{L}(p_{1})e^{2ip_{1}\ell_{13}}h_{YY}(-u,v)\notag\\
&-B_{L}(p_{2})e^{2ip_{2}\ell_{13}}h_{YY}(u,-v)+S(p_{2},p_{1})S(p_{1},-p_{2})B_{L}(p_{1})B_{L}(p_{2})e^{2i(p_{1}+p_{2})\ell_{13}}h_{YY}(-u,-v).
\end{align}

Even for the other configurations such as $C_{123}^{11\circ}$, the expression at the finite coupling can be managed in similar way. 
The rule is that {\it the structure constant is given by the summation over the sign flipping hexagon form factor with negative sign and appropriate factors related dynamical processes of the magnons}.

Since the hexagon form factor itself was determined at finite coupling by symmetry and an integrable bootstrap method \cite{BKV}, 
the spin-chain analysis should be reproduced by considering a weak coupling expansion of the hexagon form factor.
In our setup, the inserted operators are interpreted as three open spin-chains with excitations which are realized as magnons on physical edges of the hexagon.
Surely, to manage the boundary information of the hexagon is a nontrivial part. 
However, we could actually check that there is no new object for the bootstrap method at least up to $SU(2)$ sectors and the leading order of the weak coupling.\footnote{Honestly it does not guarantee that such a new object is always unnecessary for integrable bootstrap of open string correlators. For example, the higher rank sector of the Wilson loops with operator insertions or other integrable open strings such as maximal giant gravitons may have some nontrivial contributions from the boundary. To clarify this point would definitely be an invaluable future direction.}
Furthermore, since the exact reflection matrix ${\mathbb R}(p)$ is already available from \cite{Drukker:2012de, Correa:2012hh},\footnote{ 
At the weak coupling of $SU(2)$ sector, ${\mathbb R}(p)$ just becomes a phase $R(p)=e^{i p}$. As we shall analyze later, our reflection phase $B_{L}(p)$ is not the conventional reflection phase $R(p)$ since we shall use the half-step shifted basis \cite{OTY}. In usual, the one-magnon wave function in an open-chain is written as $ \psi^{(1)}(x) = \alpha(p) e^{i p x} + \alpha'(p) e^{-i p x}$. The conventional left reflection phase is defined by $R_{\rm left}(p)\equiv \frac{\alpha'(p)}{\alpha(p)}$. Then, we have $ \psi^{(1)}(x) = \alpha(p) \left( e^{i p x} + R_{\rm left}(p) e^{-i p x} \right)$. On the other hand, our boundary $S$-matrix $B_{L}(p)$ is related to the reflection matrix $R_{\rm left}(p)$ as $R_{\rm left}(p) = B_{L}(p) e^{ip} e^{2ipL}$. Thus, the reflection phase corresponding to the $SU(2)$ part of the exact reflection matrix in \cite{Drukker:2012de, Correa:2012hh} becomes the Neumann boundary condition which is realized with $B_{L}(p)= 1$.} the generalization can be straightforwardly performed.

In next section, we shall compute structure constants at tree level of the $SU(2)$ sector. 
Unlike at higher-loops or in full sectors, we need not to consider the nested procedure through the nondiagonal reflection matrix with the nontrivial dressing phase and we can understand their dynamical information well.
If we now consider the weak coupling limit of (\ref{normmulti}), the difference is just the coupling dependence.
Namely the $S$-matrix, the Bethe-Yang equation, the reflection matrix  and the hexagon form factor should be replaced by their weak coupling forms.
We shall explicitly check this by computing tree-level structure constants through open chain Bethe wavefunctions.

\section{Tree-level analysis for the $SU(2)$ sector \label{sec3}}
In previous section, we gave a conjecture for asymptotically exact structure constants of local operators on Wilson loops. 
At the leading order of weak coupling regime, it reduces to the tree-level three-point functions from open spin-chains. 
To check it, we shall explicitly calculate the correlation functions of three operators corresponding to open spin-chains at the tree-level of the $SU(2)$ sector. 
Since the scalar excitations in this sector are not overlapped with the scalar in the Wilson loops, 
one can just consider contractions between the excitation fields in the inserted operators realized by open spin chains.

\subsection{Coordinate Bethe ansatz for open spin-chain \label{sec31}}
According to \cite{MZ}, the one-loop anomalous dimension of the planar ${\cal N}=4$ super Yang-Mills corresponds to an integrable spin-chain Hamiltonian, and the spin-chain state is given by an eigenfunction obtained by diagonalizing the Hamiltonian.
The one-loop anomalous dimension for our setup is given as the eigenvalue of the $XXX_{\frac{1}{2}}$ open spin-chain Hamiltonian with Neumann boundary conditions \cite{DK}. Let us briefly review some important results in \cite{OTY} before our computations start.\footnote{We shall use slightly different conventions to \cite{OTY}. In our notation, processes of the magnons on ``coordinate axes'' become more clear. Thus, when we compare the tree level results to our conjecture, it would be better to use. The relation in detail is introduced in appendix \ref{appa}.} 

For general boundary conditions, the open spin-chain Hamiltonian appearing in the Wilson loops can be expressed as, 
\begin{align}
{\cal H}=\sum_{k=1}^{L-1}(I_{k,k+1}-P_{k,k+1})+C_{1}(I-Q_{1}^{\phi_{6}})+C_{L}(I-Q^{\phi_{6}}_{L}),
\end{align}
where $I_{k,k+1}$ is the identity operator in flavor-space and $P_{k,k+1}$ is the permutation operator which switches two scalars at site $k$ and at site $k+1$ each other. Here, the operators $Q_{1}^{\phi_{6}}$ and $Q_{L}^{\phi_{6}}$ appeared since we choose the scalar in the Wilson loop as $\phi_{6}$. As a result, $Q_{1}^{\phi_{6}}$ and $Q_{L}^{\phi_{6}}$ are defined as\footnote{The notation was introduced in \cite{Berenstein:2005vf}. In addition, although we shall not introduce the computations, we actually have calculated the $SO(6)$ Hamiltonian by evaluating all Feynman diagrams, and could check that the result is written as the above form. }
\begin{align}
Q_{1}^{\phi_{6}}|\phi_{6}\cdots\rangle=0,\hspace{1cm}Q_{1}^{\phi_{6}}|Z\cdots\rangle=|Z\cdots\rangle,\notag\\
Q_{L}^{\phi_{6}}|\cdots \phi_{6}\rangle=0,\hspace{1cm}Q_{L}^{\phi_{6}}|\cdots Z\rangle=|\cdots Z\rangle.
\end{align}
The coefficients $C_{1}$ and $C_{L}$ determine the boundary condition of the Bethe wave function. 
For example, the boundary coefficients for the $SU(2)$ sector are given as $C_{1}=C_{L}=0$ \cite{DK} and the wave function satisfies the Neumann boundary condition. Notice that for the higher rank sectors such as the $SO(6)$ sector, the boundary coefficients $C_{1}$ and $C_{L}$ become non-zero.\footnote{For the $SO(6)$ sector, the diagrams directly contracting scalar excitations with the Wilson loop itself should be included even at the one-loop level. Thus, the contribution from such diagrams will give non-zero values to the boundary terms $C_{1}$ and $C_{L}$.}

\begin{figure}[t]
\begin{center}
\includegraphics[width=15cm]{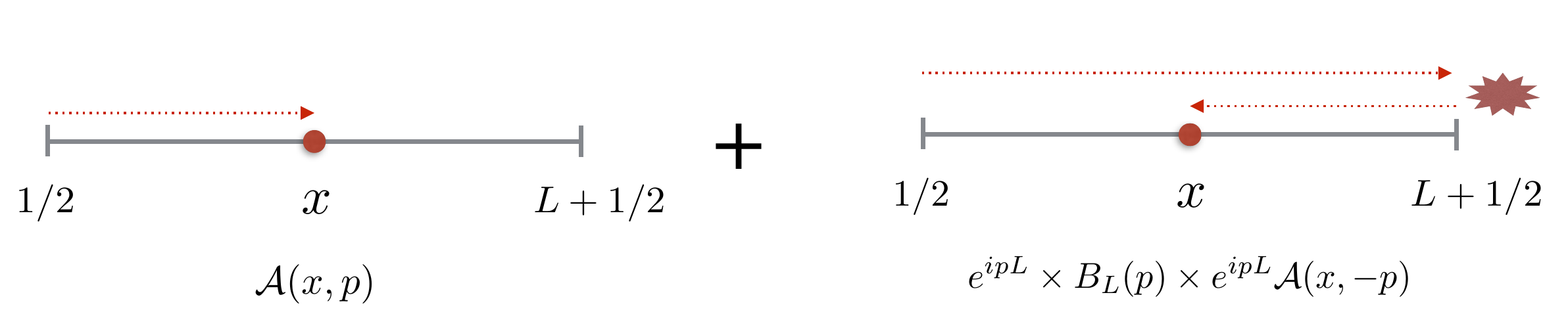}
\caption{The first term describes the magnon moving on the open spin-chain from the point at the half-step shifted site (the left boundary) to the site $x$ with the propagation factor ${\cal A}(x,p)$. On the other hand, the second term means that the magnon first moves to the right boundary with the factor $e^{ipL}$. The boundary factor $B_{L}(p)$ appears when the magnon with momentum $p$ reflects at the boundary site $L+\frac{1}{2}$. 
At the same time the magnon momentum has a sign flip as $p\rightarrow -p$. Finally, the magnon with $-p$ moves to the site $x$ by the propagation factor $e^{ipL}{\cal A}(x,-p)=e^{ip(L-(x-\frac{1}{2}))}$.}
\label{fig:1wave}
\end{center}
\end{figure}

Let us begin by giving the explicit form of the open spin-chain state for a few number of magnons in our conventions. The eigenfunction for one-magnon $\mid \Psi^{(1)}\rangle$ is defined as
\begin{align} 
{\cal H}|\Psi^{(1)}\rangle=E^{(1)}|\Psi^{(1)}\rangle. \label{eigen1}
\end{align}
Then, the Bethe state is written as
\begin{align}
&|\Psi^{(1)}\rangle=\sum_{1\leq x\leq L}\psi^{(1)}(x)|Z\cdots Z \stackrel{\stackrel{x}{\downarrow}}{Y} Z\cdots Z\rangle,\notag\\
&\psi^{(1)}(x)=A'(p)\Big({\cal A}(x,p)+e^{2ipL}B_{L}(p){\cal A}(x,-p)\Big)\ \,\ {\rm with} \,\ \ B_{L}(p)=-\frac{e^{-ip}-(1-C_{L})}{1-(1-C_{L})e^{-ip}},\label{1wave}
\end{align}
where $A'(p)$ is a normalization factor and ${\cal A}(x,p)\equiv e^{ip(x-\frac{1}{2})}$ is the propagation factor shifted by the half-step.\footnote{The half-step shift has been first introduced in \cite{DeWolfe:2004zt}.} 
It is straightforward to understand the dynamical processes of the magnon on the coordinate axis from the wave function (\ref{1wave}), see also figure \ref{fig:1wave}.
The details of the wave function are analyzed in appendix \ref{appa}.

\begin{figure}[t]
\begin{center}
\includegraphics[width=15cm]{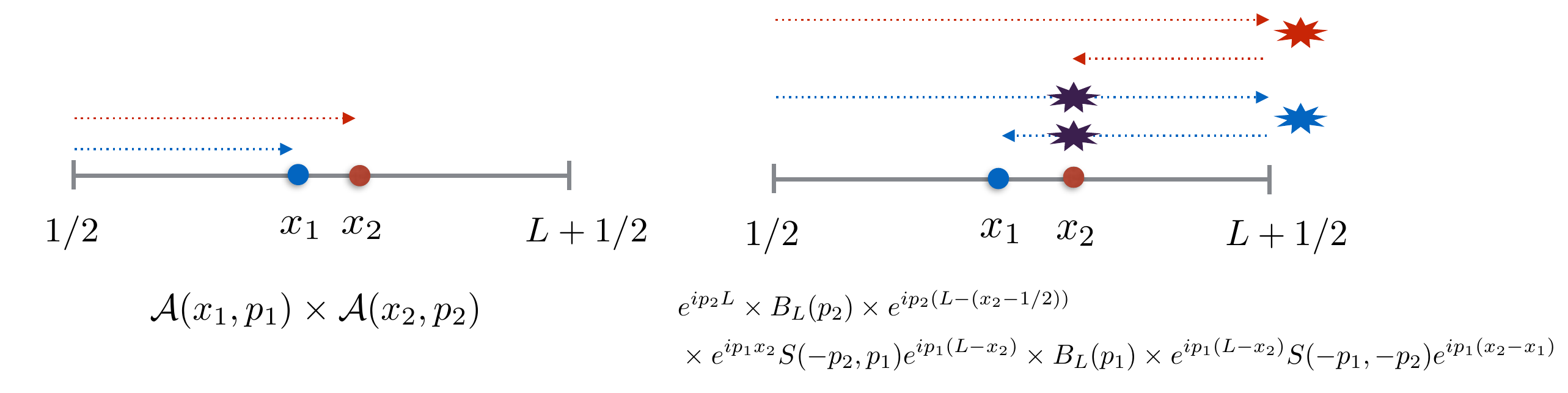}
\caption{For the first term ${\cal A}(x_{1}, p_{1}){\cal A}(x_{2}, p_{2})$, two magnons moves to their positions $x_{1,2}$ independently without any nontrivial factors. On the other hand, the term $S(p_{2},p_{1})S(-p_{2},p_{1}) e^{2i(p_{1}+p_{2})L}B_{L}(p_{1})B_{L}(p_{2}){\cal A}(x_{1},-p_{1}){\cal A}(x_{2},-p_{2})$ is non-trivial since the magnons scatter off from each other. Firstly, the magnon with momentum $p_{2}$ reaches at the site $x_{2}$ through the factor $e^{ip_{2}}\times B_{L}(p_{2})\times e^{ip_{2}(L-(x_{2}-\frac{1}{2}))}$. This process is the same as the second term of the one-magnon wave function. Next, we move the magnon with momentum $p_{1}$. Then, the $S$-matrix factors are necessary as the magnons scatters.}
\label{fig:2wave}
\end{center}
\end{figure}

We next consider the two-magnon case whose eigenfunction is defined as 
\begin{align} 
{\cal H}|\Psi^{(2)}\rangle=E^{(2)}|\Psi^{(2)}\rangle\label{eigen2}.
\end{align}
The Bethe ansatz state for the two-magnon is
\begin{align}
|\Psi^{(2)}\rangle &=\sum_{1\leq x_{1}< x_{2}\leq L}\psi^{(2)}(x_{1},x_{2})|Z\cdots Z \stackrel{\stackrel{x_{1}}{\downarrow}}{Y} Z\cdots Z \stackrel{\stackrel{x_{2}}{\downarrow}}{Y} Z\rangle\notag ,\\
\psi^{(2)}(x_{1},x_{2})/A'(p_{1},p_{2})&=f(x_{1},p_{1};x_{2},p_{2})+e^{2ip_{2}L}B_{L}(p_{2})f(x_{1},p_{1};x_{2},-p_{2})\notag\\
&+S(p_{2},p_{1})S(-p_{2},p_{1})e^{2ip_{1}L}B_{L}(p_{1})f(x_{1},-p_{1};x_{2},p_{2})\label{2wave}\\
&+S(p_{2},p_{1})S(-p_{2},p_{1})e^{2i(p_{1}+p_{2})L}B_{L}(p_{1})B_{L}(p_{2})f(x_{1},-p_{1};x_{2},-p_{2}),\notag
\end{align}
where $A'(p_{1},p_{2})$ is a normalization factor. Note that the $S$-matrix for $SU(2)$ sector is given as $S(p_{2},p_{1})=\frac{u-v-i}{u-v+i}$. Thus the S-matrix has the nice property : $S(-p_{i},p_{j})=S(-p_{j},p_{i})$. We used the notation for rapidity variables $u=\cot\frac{p_{1}}{2}$ and $v=\cot\frac{p_{2}}{2}$.  
Here we defined the factor $f(x_{1},p_{1};x_{2},p_{2})$  as
\begin{align}
f(x_{1},p_{1};x_{2},p_{2})\equiv{\cal A}(x_{1},p_{1}){\cal A}(x_{2},p_{2})+S(p_{2}, p_{1}){\cal A}(x_{1},p_{2}){\cal A}(x_{2},p_{1}),\label{2f}
\end{align}
which is interpreted as the terms given by the summation over the permutations. The dynamical processes of the two-magnon wave function are understood as in figure \ref{fig:2wave}.\footnote{For intuitive understanding, it may be better to rewrite the S-matrix factor $S(p_{2},p_{1})S(-p_{2},p_{1})$ as $S(p_{2},p_{1})S(-p_{1},p_{2})$.} 
Notice that it is easily found that the wave function is basically constructed by the summation over the sign flipping terms of $f(x_{1},p_{1};x_{2},p_{2})$, that is, 
\begin{align} 
f(x_{1},p_{1};x_{2},p_{2}),\ f(x_{1},-p_{1};x_{2},p_{2}),\ f(x_{1},p_{1};x_{2},-p_{2}),\ f(x_{1},-p_{1};x_{2},-p_{2}).\label{allpassible}
\end{align}
However, because of the scatterings of the magnons, we must insert the matrices factors times propagation factors such as $e^{2ip_{1}L}B_{L}(p_{1})S(p_{2},p_{1})S(-p_{2},p_{1})$. For the common point, when the sign of the momentum $p_l$ is flipped, we must insert the $S$-matrix factor 
\begin{align}
e^{2ip_{l}L}B_{L}(p_{l})\prod_{k>l}S(p_{k},p_{l})S(-p_{k},p_{l}).\label{2wavefactor}
\end{align}
The factors depend on dynamical situations.\footnote{For example, let us consider $M$-magnon problem. After choosing the leftmost magnon with the momentum $p_{1}$, let us think of how to obtain the sign flipped momentum of the magnon. If the magnon reflect at the right boundary and comes back to the original position, we have to put a factor in the wave function such as $S(p_{1}, p_{2})\cdots S(p_{1}, p_{M})R^{\rm Right}(p_{1})S(p_{M}, -p_{1})\cdots S(p_{2}, -p_{1})$. In general, we have to consider all possible dynamical situations with appropriate factors.} Based on these systematic constructions, we can find the multi-magnon open spin chain wave function.

Similarly, the $M$-magnon wave function can be written down:
\begin{align}
\psi^{(M)}&=\underbrace{\sum_{{\sf P}_{+}\cup {\sf P_-}=\{1,\ldots,M\}}\left[\prod_{l\in {\sf P}_{-}}(e^{2i p_l L})\prod_{k>l}S(p_k,p_l)S(-p_k,p_l)\right]}_{\rm summation\ over\ the\ sign\ flipping\ terms\ with\ appropriate\ factors}f(\hat{p}_{1},\cdots,\hat{p}_{M}), \cr
&f(\hat{p}_{1},\cdots,\hat{p}_{M})\equiv\underbrace{\sum_{\sigma_{1}\neq \cdots\neq\sigma_{M}}^{M}\prod_{\underset{\sigma_{k}<\sigma_{j}}{j<k}}S(\hat{p}_{\sigma_{j}},\hat{p}_{\sigma_{k}})}_{\rm summation\ over\ the\ permutation}\prod_{m=1}^{M}{\cal A}(x_{m},\hat{p}_{\sigma_{m}}),\label{multi}
\end{align}
where $\hat{p}_{i}$ is defined as
\begin{align}
\hat{p}_i=\begin{cases}p_i \qquad & i\in {\sf P}_{+}\\-p_i \qquad & i\in {\sf P}_{-}\end{cases}.
\end{align}
The $M$-magnon wave function is constructed by the two parts : The summation over the sign flipping terms with appropriate factors related to process of the magnon which is written in the first line, and in second line the summation over the permutation part.

\subsection{Structure constants and the hexagon form factor $h_{YY}(u,v)$ \label{sec32}}
Using the open spin chain wave functions discussed above, we now calculate the structure constants at tree-level. 
Thereby we would like to express the structure constants in terms of the hexagon form factor $h_{YY}(u,v)$.

\subsubsection{A nontrivial operator with one-magnon : $C^{1\circ\circ}_{123}$ \label{sec321}}

\begin{figure}[t]
\begin{center}
\includegraphics[width=5cm]{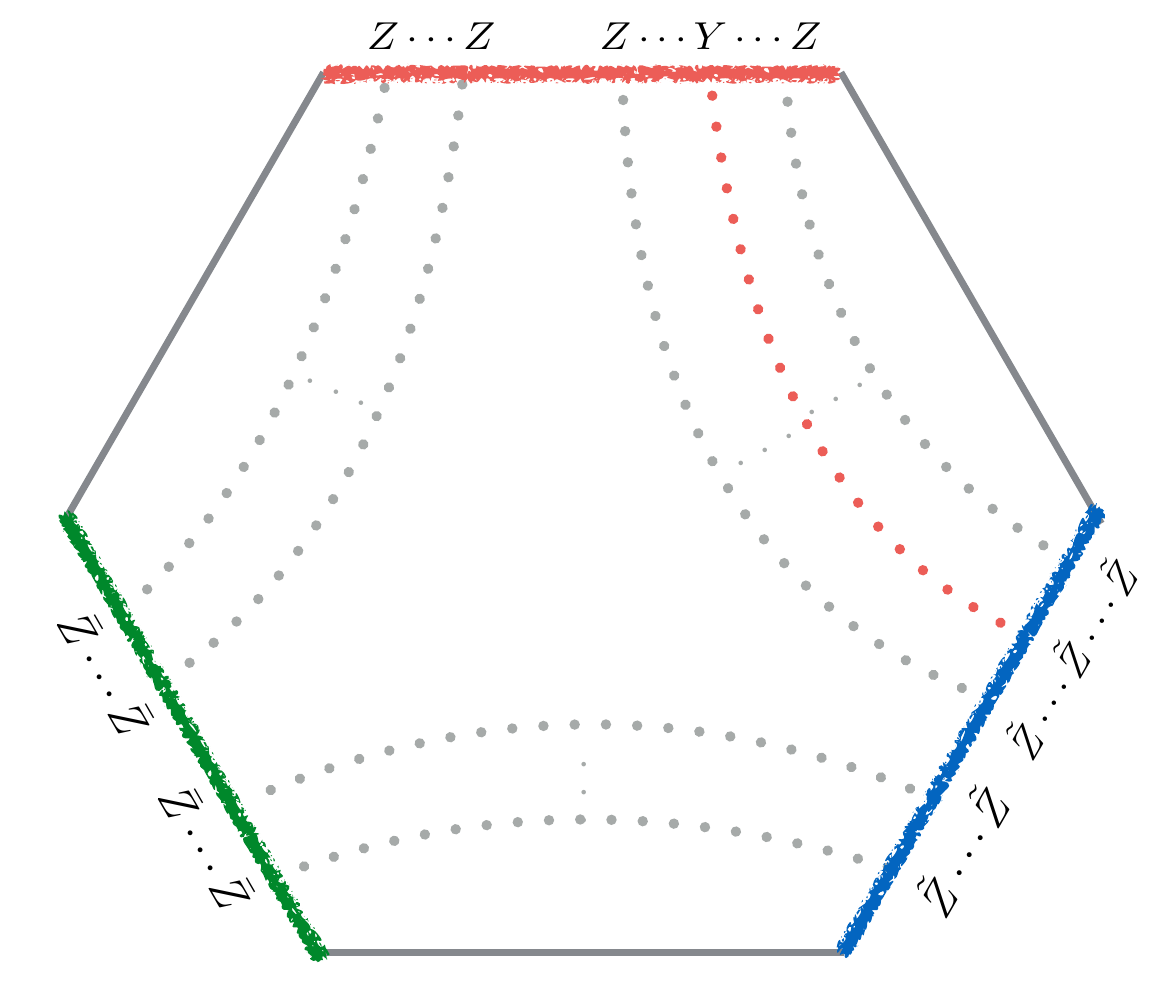}
\caption{For Y-excitation on the top spin chain case, the magnon can be contracted with only the vacuum $\tilde{Z}$. Therefore the structure constant $C_{123}^{1\circ\circ}$ is given by the summation over the positions of the 1-magnon wave function.}
\label{fig:1}
\end{center}
\end{figure}

\begin{figure}[t]
\begin{center}
\includegraphics[width=10cm]{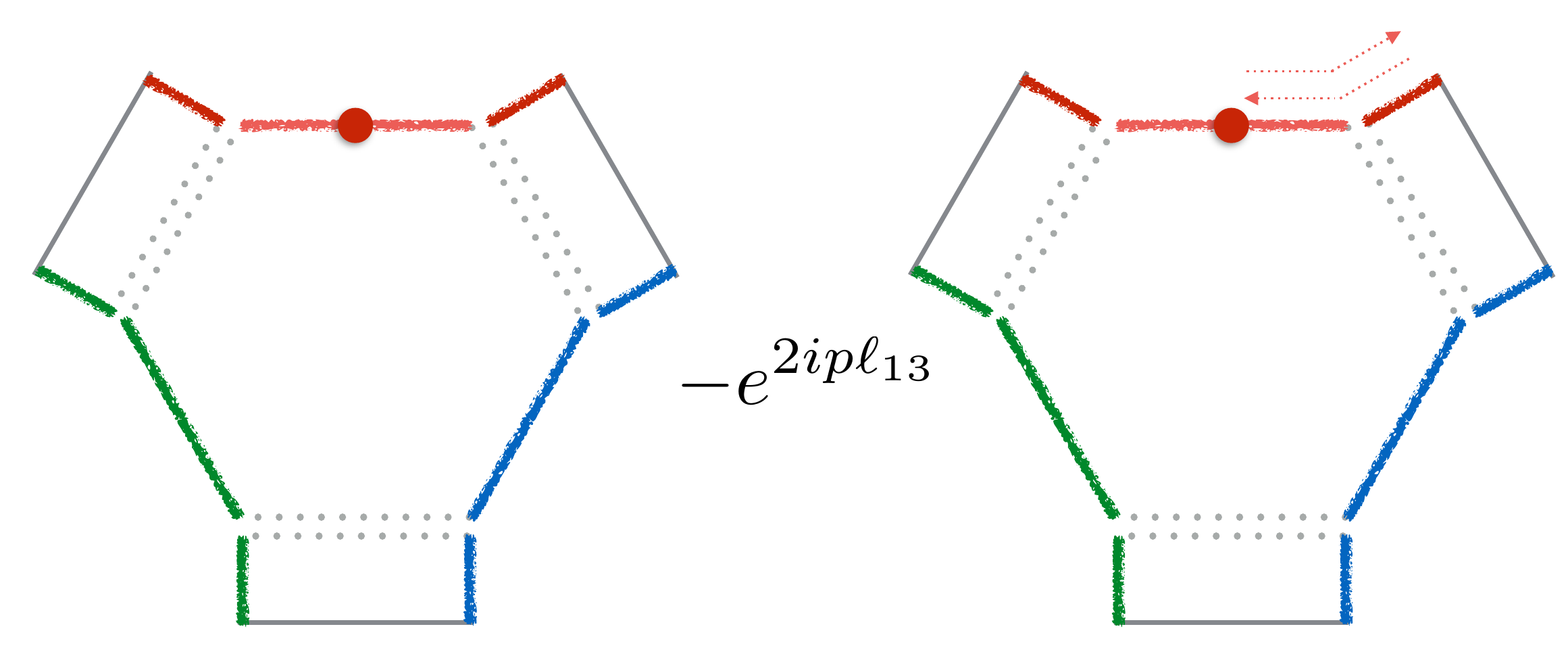}
\caption{The first term is just represented as a magnon on the hexagon. On the other hand, the second term have the propagation factor $e^{2ip\ell_{13}}$. This means that firstly a magnon with momentum $p$ move to the left boundary and is reflected at the boundary. Next, the magnon with momentum $-p$ move to the splitting point by the propagation factor $e^{-ip(-\ell_{13})}$}
\label{fig:process1mag}
\end{center}
\end{figure}

The structure constants at tree-level are obtained by summing over the Wick contractions. 
Let us begin with the simplest case which is described in figure \ref{fig:1}. It is made of a non-trivial operator with an excitation and two trivial operators. We denote such a structure constant by $C_{123}^{1\circ\circ}$.
The first spin chain has a magnon as ${\cal O}_{1}:\sum_{x}Z^{x}YZ^{L_{1}-(x+1)}$ and the others are vacuum states as ${\cal O}_{2}:\bar{Z}^{L_{2}}$ and ${\cal O}:\tilde{Z}^{L_{3}}$. In this case, the structure constants are simply calculated by contraction between the first spin-chain and the third spin-chain :
\begin{align}
C_{123}^{1\circ\circ}\propto \sum_{x_{2}=\ell_{12}+1}^{L_{1}}\psi^{(1)}(x).
\end{align}
The summation of the propagator ${\cal A}(x,p)$ becomes
\begin{align}
\sum_{x_{2}=\ell_{12}+1}^{L_{1}}{\cal A}(x,p)={\cal M}(p)(e^{ip\ell_{12}}-e^{ipL_{1}}),
\end{align}
where the factor ${\cal M}(p)\equiv(e^{-i\frac{p}{2}}-e^{i\frac{p}{2}})^{-1}$ obeys a useful identity  ${\cal M}(p)=-{\cal M}(-p)$. 
Therefore, we get
\begin{align}
C_{123}^{1\circ\circ}&\propto {\cal M}(p)(e^{ip\ell_{12}}-e^{2ipL_{1}}e^{-ip\ell_{12}})\notag\\
&={\cal M}_{\ell_{12}}(p)(1-e^{2ip\ell_{13}}),\hspace{1cm}{\cal M}_{\ell_{12}}(p)\equiv {\cal M}(p)e^{ip\ell_{12}}\label{1result}
\end{align}
In this normalization due to factor out the propagation by the spin chain length $\ell_{12}$, the magnon starts at the splitting point in figure \ref{fig:1}. It is easy to understand the result as a moving dynamical process on the hexagon form factor in figure \ref{fig:process1mag}. We find that the magnon of the second term have negative sign against the magnon of the first term in figure \ref{fig:process1mag}. Thus the relative factor $e^{2ip\ell_{13}}$ can be understood as the process in order to get the magnon with negative momentum.

Let us give some comments for this result. 
First, the explicit difference from the closed spin chain case is the negative momentum term. 
The structure constant of the closed spin chain with one-magnon has only the first term in (\ref{1result}) because of absence of reflections. 
Second, every magnons are at splitting point on the hexagon form factor, not at the boundary. 
This can be understood because we are considering the $SU(2)$ sector where the boundary condition is Neumann and $B_{L}(p)=1(C_{L}=0)$ in (\ref{1wave}). 
On the other hand, if we consider the higher rank sectors with other excitations where $B_{L}(p)\neq 1$, 
the terms for a magnon living at the boundary may survive. 
This contribution may imply an additional object for bootstrap.

\subsubsection{A nontrivial operator with two-magnon : $C^{2\circ\circ}_{123}$ \label{sec322}}

\begin{figure}[t]
\begin{center}
\includegraphics[width=13cm]{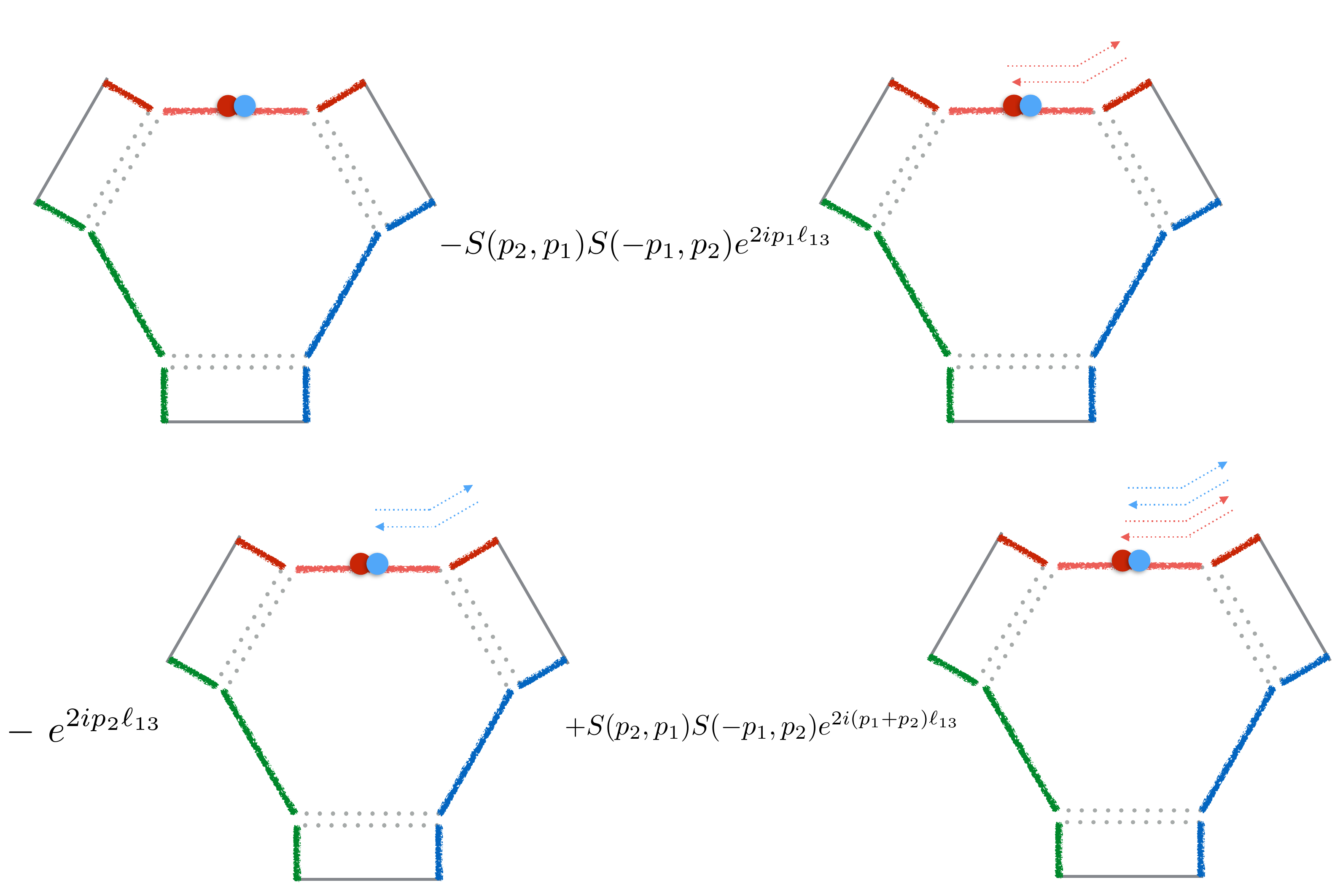}
\caption{For the second term, the red colored magnon on the hexagon can get the negative sign $-p_{1}$ by the propagation $e^{2ip_{1}\ell_{13}}$. Then, since the magnon would be scattered with another magnon with $p_{2}$ at two times, the S-matrix factors $S(p_{2},p_{1})S(-p_{1},p_{2})$ are needed. In the same way, the third and fourth terms are interpreted as the magnons with $(p_{1},-p_{2})$ and the magnons with $(-p_{1},-p_{2})$ on the hexagon  respectively by added appropriate propagation and S-matrix factors. }
\label{fig:process}
\end{center}
\end{figure}

Next we consider the case of a nontrivial operator with two excitations and two trivial operators, namely where the first spin chain is set to ${\cal O}_{1}:\sum_{x<y}Z^{x}YZ^{y-(x+1)}YZ^{L_{1}-(y+1)}$ and the others are vacuum states. We denote this by $C_{123}^{2\circ\circ}$. 
In this case, the structure constants become
\begin{align}
C_{123}^{2\circ\circ}\propto \sum_{1\leq x_{1}<x_{2}\leq L_{1}}\psi^{(2)}(x_{1},x_{2})=\sum_{x_{1}=\ell_{12}+1}^{L_{1}}\sum^{L_{1}}_{x_{2}=x_{1}+1}\psi^{(2)}(x_{1},x_{2}).\label{200}
\end{align}
Notice that we extend the summation range because the contribution of $x_{1}=L_{1}$ becomes zero. 
The summation of the product of the propagators ${\cal A}(x_{1},p_{1}){\cal A}(x_{2},p_{2})$ in $\psi^{(2)}$ is written as
\begin{align}
&\sum_{1\leq x_{1}<x_{2}\leq L_{1}}{\cal A}(x_{1},p_{1}){\cal A}(x_{2},p_{2})\notag\\
&={\cal M}(p_{1}){\cal M}(p_{2})\Big\{\frac{i+2v}{2(u+v)}e^{i(p_{1}+p_{2})\ell_{12}}-e^{ip_{2}L_{1}}e^{ip_{1}\ell_{12}}+\Big(-\frac{i+2v}{2(u+v)}+1\Big)e^{i(p_{1}+p_{2})L_{1}}\Big\}.\label{AA}
\end{align}
The factor $\frac{i+2v}{2(u+v)}$ comes from appropriately normalizing geometric series. 
Furthermore, (\ref{AA}) is constructed by the propagations over both the bridge length $e^{ip\ell_{12}}$ and the spin chain length $e^{ipL_{1}}$. 
This fact can be understood clearly as dynamical processes of the magnons on the spin chain as in appendix \ref{appb}. 
According to the coordinate Bethe ansatz for the two-magnon (\ref{2wave}), we sum over all possible processes. 
Then, surprisingly, the other terms except for the bridge length propagation term given by $e^{i(p_{1}+p_{2})\ell_{12}}$ completely vanish. 
Therefore we finally obtain
\begin{align}
C_{123}^{2\circ\circ}&\propto{\cal M}_{\ell_{12}}(p_{1}){\cal M}_{\ell_{12}}(p_{2})\Big\{\frac{u-v}{i+u-v}-S(p_{2},p_{1})S(-p_{2},p_{1})e^{2ip_{1}\ell_{13}}\frac{-u-v}{i-u-v}\notag\\
&-e^{2ip_{2}\ell_{13}}\frac{u+v}{i+u+v}+S(p_{2},p_{1})S(-p_{2},p_{1})e^{2i(p_{1}+p_{2})\ell_{13}}\frac{-u+v}{i-u+v}\Big\}.\label{2st}
\end{align}
Here, we obtained a nontrivial factor, i.e. $\frac{u-v}{i+u-v}$. Actually, this factor is known as the hexagon form factor at tree-level
\begin{align*}
h_{YY}(u,v)=\frac{u-v}{i+u-v}+{\cal O}(g).
\end{align*}
Furthermore the result of the structure constant (\ref{2st}) can be explained by the dynamical process on the hexagon form factor in figure \ref{fig:process}.

\subsubsection{A nontrivial operator with $M$-magnon : $C_{123}^{M\circ\circ}$ \label{sec323}}
By doing similar tasks, we would like to get the structure constants for the multi-magnon.
The wave function for the multi-magnon is naively written as
\begin{align}
\sum_{x_{1}<\cdots<x_{M}}\psi^{(M)}(x_{1},\cdots,x_{M}).\label{wM}
\end{align}
The summation of the multi-magnon wave function should be obtained by summing the hexagon form factor over all patterns such as flipping momentum signs with the negative weight. To justify the above statement, we prove the following two lemmas :
\begin{enumerate}
\item Multi-magnon hexagon form factor : The hexagon form factor contains only the bridge length dependent terms such as $e^{i(p_{1}+\cdots+p_{M})\ell_{12}}$. 
\item Bridge length independent terms : The others terms which are independent of the bridge length vanish.
\end{enumerate}

{\bf 1. Multi-magnon hexagon form factor}\\

Let us first explain how to generalize the few number of magnons result to a multi-magnon result.\footnote{One can apply the same argument to the computation of the multi-magnon form factors in the ordinary gauge invariant operators; see \cite{Les}. } We focus on the summation over the permutation parts of the multi-magnon wave function (\ref{multi}):
\begin{align}
\sum_{x_{1}<\cdots<x_{M}}\sum_{\sigma_{1}\neq\cdots\neq \sigma_{M}}\prod_{\overset{\sigma_{k}<\sigma_{j}}{j<k}}S(p_{\sigma_{j}},p_{\sigma_{k}}){\cal A}(x_{1},p_{\sigma_{1}})\cdots{\cal A}(x_{M},p_{\sigma_{M}}).\label{hex}
\end{align}
As the above summation for a few number of magnons is relatively manageable, it is given in appendix \ref{appc}.
On the other hand, the corresponding multi-magnon summation is a little more complicated.
We start with the fact that the $S$-matrix is related to the hexagon form factors by
\begin{align}
S(u,v)=\frac{h(v,u)}{h(u,v)}.
\end{align}
By using this, the product of $S$-matrices can be transformed as
\begin{align}
\prod_{\overset{\sigma_{k}<\sigma_{j}}{j<k}}S(p_{\sigma_{j}},p_{\sigma_{k}})=\left(\prod_{\overset{\sigma_{k}<\sigma_{j}}{j<k}}h(u_{\sigma_{k}},u_{\sigma_{j}})\right)\left(\prod_{\overset{\sigma_{k}<\sigma_{j}}{j<k}}\frac{1}{h(u_{\sigma_{j}},u_{\sigma_{k}})}\right).
\end{align}
Furthermore, the first bracket can be decomposed as
\begin{align}
\left(\prod_{\overset{\sigma_{k}<\sigma_{j}}{j<k}}h(u_{\sigma_{k}},u_{\sigma_{j}})\right)=\left(\prod_{\sigma_{k}<\sigma_{j}}h(u_{\sigma_{k}},u_{\sigma_{j}})\right)\left(\prod_{\overset{\sigma_{k}<\sigma_{j}}{j>k}}\frac{1}{h(u_{\sigma_{k}},u_{\sigma_{j}})}\right).
\end{align}
Thus, by relabelling indices of the product, we get
\begin{align}
\prod_{\overset{\sigma_{k}<\sigma_{j}}{j<k}}S(u_{\sigma_{j}},u_{\sigma_{k}})=\left(\prod_{j<k}h(u_{j},u_{k})\right)\left(\prod_{\overset{\sigma_{k}>\sigma_{j}}{j<k}}\frac{1}{h(u_{\sigma_{k}},u_{\sigma_{j}})}\right)\left(\prod_{\overset{\sigma_{k}<\sigma_{j}}{j<k}}\frac{1}{h(u_{\sigma_{k}},u_{\sigma_{j}})}\right).
\end{align}
In addition, we have
\begin{align}
\prod_{\overset{\sigma_{k}<\sigma_{j}}{j<k}}S(u_{\sigma_{j}},u_{\sigma_{k}})=\left(\prod_{j<k}h(u_{j},u_{k})\right)\left(\prod_{j<k}\frac{1}{h(u_{\sigma_{j}},u_{\sigma_{k}})}\right).
\end{align}
Therefore, the summation (\ref{hex}) is rewritten as
\begin{align}
\left(\prod_{j<k}h(u_{j},u_{k})\right)\sum_{\sigma_{1}\neq\cdots\neq \sigma_{M}}\left(\prod_{j<k}\frac{1}{h(u_{\sigma_{j}},u_{\sigma_{k}})}\right)\sum_{x_{1}<\cdots<x_{M}}{\cal A}(x_{1},p_{\sigma_{1}})\cdots{\cal A}(x_{M},p_{\sigma_{M}}).
\end{align}
The first bracket is just the multi-magnon hexagon form factor since it can be decomposed by the two-magnon hexagon form factor
\begin{align}
h(u_{1},\cdots,u_{M})=\prod_{j<k}h(u_{j},u_{k}).
\end{align}

\begin{figure}[t]
\begin{center}
\includegraphics[width=15cm]{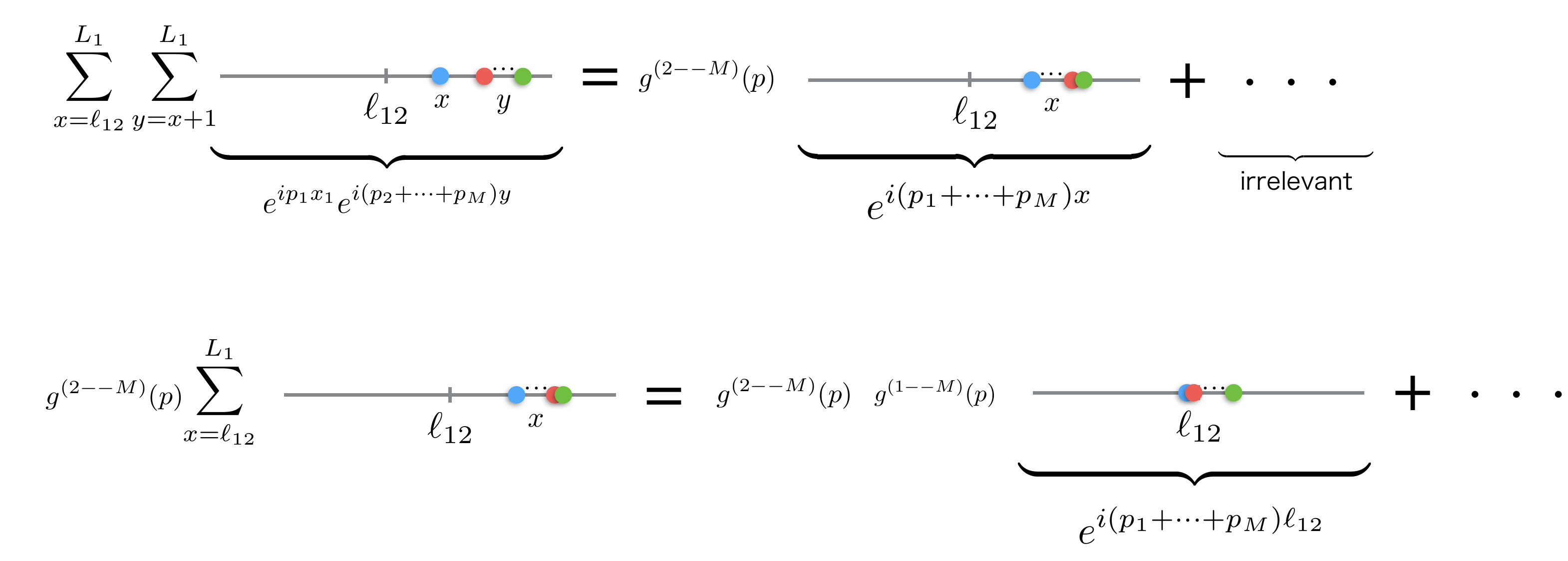}
\caption{We suppose that a magnon is located at the site $x$ and the remaining $M-1$ magnons are located at the site $y$. Firstly, we sum over the position $y$. Then we obtain a geometric series such as $g^{(2--M)}=\frac{1}{e^{-i(p_{2}+\cdots+p_{M})}-1}$, and have the propagation factor $e^{i(p_{1}+\cdots+p_{M}x)}$. Next, by summing over the position $x$, we get the factor $g^{(1--M)}=\frac{1}{e^{-i(p_{1}+\cdots+p_{M})}-1}$. Therefore, the coefficient in front of the bridge length propagation factor $e^{i(p_{1}+\cdots+p_{M})}$ becomes the product for the geometric series.}
\label{fig:geometric}
\end{center}
\end{figure}

Let us next treat the bridge length dependent parts in the summation over the positions given as
\begin{align}
\sum_{x_{1}<\cdots<x_{M}}{\cal A}(x_{1},p_{\sigma_{1}})\cdots{\cal A}(x_{M},p_{\sigma_{M}}),
\end{align}
which is easily evaluated by geometric series. 
For one- and two-magnon cases, they respectively become
\begin{align*}
\sum_{x}{\cal A}(x,p)&=e^{-\frac{i}{2}p}\frac{1}{e^{-ip}-1}e^{ip\ell_{12}}+\cdots,\\
\sum_{x_{1}<x_{2}}{\cal A}(x_{1},p_{1}){\cal A}(x_{2},p_{2})&=e^{-\frac{i}{2}(p_{1}+p_{2})}\frac{1}{e^{-ip_{2}}-1}\frac{1}{e^{-i(p_{2}+p_{1})}-1}e^{i(p_{1}+p_{2})\ell_{12}}+\cdots .
\end{align*}
Then, the summation for multi-magnon would become
\begin{align}
\sum_{x_{1}<\cdots<x_{M}}{\cal A}(x_{1},p_{1})\cdots{\cal A}(x_{M},p_{M})
=\prod_{k=1}^{M}e^{-\frac{i}{2}p_{k}}\prod_{j=1}^{M}\frac{1}{e^{-i\sum_{k=j}^{M}p_{k}}-1}e^{i(p_{1}+\cdots+p_{M})\ell_{12}}+\cdots .
\end{align}
The geometric series in the summation is sketched in figure \ref{fig:geometric}.\footnote{See also appendix \ref{appb}.}

From above argument the summation (\ref{hex}) becomes
\begin{align}
\prod_{k=1}^{M}e^{-\frac{i}{2}p_{k}}\left(\prod_{s<t}h(u_{s},u_{t})\right)\sum_{\sigma_{1}\neq\cdots\neq \sigma_{M}}\left(\prod_{j<k}\frac{1}{h(u_{\sigma_{j}},u_{\sigma_{k}})}\right)\prod_{n=1}^{M}\frac{1}{e^{-i\sum_{m=n}^{M}p_{m}}-1}e^{i(p_{1}+\cdots+p_{M})\ell_{12}}+\cdots .
\end{align}
We finally prove the following relations : 
\begin{align}
F(p_{1},\cdots,p_{M})&=\prod_{k=1}^{M}i\left(u_{k}+\frac{i}{2}\right),\\
F(p_{1},\cdots,p_{M})&\equiv \sum_{\sigma_{1}\neq\cdots\neq \sigma_{M}}\left(\prod_{j<k}\frac{1}{h(u_{\sigma_{j}},u_{\sigma_{k}})}\right)\prod_{n=1}^{M}\frac{1}{e^{-i\sum_{m=n}^{M}p_{m}}-1},\label{lemma}
\end{align}
by mathematical induction. 
First, relating $F(p_{1})$ for the $M=1$ case is trivially found. 
Next, we assume that the relation holds for $M-1$ case. Then, we extract the expression $F(p_{1},\cdots ,\check{p_{j}},\cdots,p_{M})$ where the $j$-th excitation does not contribute :
\begin{align}
F(p_{1},\cdots,p_{M})=\frac{1}{e^{-i\sum_{k=1}^{M}p_{k}}-1}\sum_{j=1}^{M}\left(\prod_{k\neq j}\frac{1}{h(u_{k},u_{j})}\right)F(p_{1},\cdots ,\check{p_{j}},\cdots,p_{M}).
\end{align}
We finally have
\begin{align}
F(p_{1},\cdots,p_{M})=\frac{1}{e^{-i\sum_{k=1}^{M}p_{k}}-1}\sum_{j=1}^{M}\left(\prod_{k\neq j}\frac{1}{h(u_{k},u_{j})}\right)\frac{\prod_{k=1}^{M}i(u_{k}+1/2)}{i(u_{j}+1/2)}.
\end{align}
Here, the expression   
\begin{align}
\frac{1}{e^{-i\sum_{k=1}^{M}p_{k}}-1}\sum_{j=1}^{M}\left(\prod_{k\neq j}\frac{1}{h(u_{k},u_{j})}\right)\frac{1}{i(u_{j}+1/2)}
\end{align}
can be handled by introducing a residue integral which is given as
\begin{align}
\oint \frac{dz}{2\pi i}\frac{1}{z}\left(\prod_{k=1}^{M}\frac{u_{k}-z-i/2}{u_{k}-z+i/2}-1\right),
\end{align}
where the integrand has poles at $z=0$ and $z=u_{k}+i/2$. Picking up the pole at $z=0$, we get 
\begin{align}
\oint_{z=0} \frac{dz}{2\pi i}\frac{1}{z}\left(\prod_{k=1}^{M}\frac{u_{k}-z-i/2}{u_{k}-z+i/2}-1\right)=e^{-i\sum_{k=1}^{M}p_{k}}-1 .
\end{align}
Otherwise, from the other poles, we have
\begin{align}
\oint_{\rm z=u_{k}+i/2} \frac{dz}{2\pi i}\frac{1}{z}\left(\prod_{k=1}^{M}\frac{u_{k}-z-i/2}{u_{k}-z+i/2}-1\right)=\sum_{j=1}^{M}\left(\prod_{k\neq j}\frac{1}{h(u_{k},u_{j})}\right)\frac{1}{i(u_{j}+1/2)}.
\end{align}
From those, we could completely get the following relation :
\begin{align}
F(p_{1},\cdots,p_{M})&=\prod_{k=1}^{M}i\left(u_{k}+\frac{i}{2}\right)\\
&=\prod_{k=1}^{M}\frac{1}{e^{-ip_{k}}-1}.
\end{align}
Thus, the summation (\ref{hex}) becomes
\begin{align}
&\sum_{x_{1}<\cdots<x_{M}}\sum_{\sigma_{1}\neq\cdots\neq \sigma_{M}}\prod_{\overset{\sigma_{k}<\sigma_{j}}{j<k}}S(p_{\sigma_{j}},p_{\sigma_{k}}){\cal A}(x_{1},p_{\sigma_{1}})\cdots{\cal A}(x_{M},p_{\sigma_{M}})\\
&=\prod_{k=1}^{M}{\cal M}(p_{k})\prod_{j<k}h(u_{j},u_{k})e^{i(p_{1}+\cdots+p_{M})\ell_{12}}+\cdots .
\end{align}

{\bf 2. Bridge length independent terms}\\

For a naive discussion, of course, we can take the bridge length to zero such as $\ell_{12}=0$.\footnote{The others are not zero such as $\ell_{13}\neq0$ and $\ell_{23}\neq0$.} Then the structure constants don't have any non-trivial factors because the Bethe state can't contract with other states. This means that the structure constants are independent for the spin chain length $L_{i}$. Therefore any terms except the bridge length dependent terms $e^{i(p_{1}+\cdots +p_{M})\ell_{12}}$ must vanish. We shall give more rigorous certification.

Now we focus on the spin-chain length dependent terms such as $e^{i(p_{1}+\cdots+p_{M})L_{1}}$.\footnote{A few number of magnon case is analyzed in appendix \ref{appd} for helping to understand.} By applying the result of the multi-magnon hexagon form factor, the summation (\ref{wM}) for the wave function  with Neumann boundary condition is given by:
\begin{align}
&\sum_{x_{1}<\cdots<x_{M}}\psi^{(M)}(x_{1},\cdots,x_{M})/\prod_{i=1}^{M}{\cal M}(p_{i})\notag\\
&=\sum_{{\sf P}_{+}\cup {\sf P_-}=\{1,\ldots,M\}}\left[\prod_{k\in {\sf P}_{-}}(-e^{2i p_k L})\prod_{l<k}S(p_k,p_l)S(-p_k,p_l)\right]\prod_{i<j}h(\hat{p}_i,\hat{p}_j)e^{i(\hat{p}_{1}+\cdots+\hat{p}_{M})L_{1}}+\cdots . \label{no}
\end{align}
First of all, the propagation factor can be trivially picked out:
\begin{align}
e^{i(p_{1}+\cdots+p_{M})L}\sum_{{\sf P}_{+}\cup {\sf P_-}=\{1,\ldots,M\}}\left[\prod_{k\in {\sf P}_{-}}(-)\prod_{l<k}S(p_k,p_l)S(-p_k,p_l)\right]\prod_{i<j}h(\hat{p}_i,\hat{p}_j)+\cdots .\label{negation}
\end{align}
By dividing the factor such as
\begin{align}
\prod_{l<k}h(p_{l},p_{k})\prod_{l<k}h(p_{l},-p_{k}),
\end{align}
the leading term $(p_{1},\cdots,p_{M})$ becomes
\begin{align}
&e^{i(p_{1}+\cdots+p_{M})L}\prod_{l<k}h(p_{l},p_{k})\prod_{l<k}h(p_{l},-p_{k})\left(\frac{1}{\prod_{l<k}h(p_{l},p_{k})\prod_{l<k}h(p_{l},-p_{k})}\prod_{i<j}h(p_i,p_j)\right)\notag\\
&=e^{i(p_{1}+\cdots+p_{M})L}\prod_{l<k}h(p_{l},p_{k})\prod_{l<k}h(p_{l},-p_{k})\left(\prod_{l<k}\frac{1}{h(p_{l},-p_{k})}\right). \label{plus}
\end{align}
On the other hand, the next to leading term which is the term for $p_{1}\rightarrow-p_{1}$ is written as
\begin{align}
&-e^{i(p_{1}+\cdots+p_{M})L}\prod_{l<k}h(p_{l},p_{k})\prod_{l<k}h(p_{l},-p_{k})\notag\\
&\times\left(\frac{1}{\prod_{l<k}h(p_{l},p_{k})\prod_{l<k}h(p_{l},-p_{k})}\prod_{l<k}^{M}S(p_{k},p_{l})S(-p_{k},p_{l})\prod_{i=1}^{M}h(-p_{1},p_{i})\prod_{1\neq i<j}^{M}h(p_{i},p_{j})\right)\notag\\
&=-e^{i(p_{1}+\cdots+p_{M})L}\prod_{l<k}h(p_{l},p_{k})\prod_{l<k}h(p_{l},-p_{k})\left(\prod_{k}\frac{1}{h(p_{1},p_{k})}\prod_{1\neq l<k}\frac{1}{h(p_{l},-p_{k})}\right),
\end{align}
where the expression in brackets is just the negative $k$ part of (\ref{plus}). Generally, the equation (\ref{negation}) becomes
\begin{align}
e^{i(p_{1}+\cdots+p_{M})L}\prod_{l<k}h(p_{l},p_{k})\prod_{l<k}h(p_{l},-p_{k})\sum_{{\sf P}_{+}\cup {\sf P_-}=\{1,\ldots,M\}}\left[\prod_{k\in {\sf P}_{-}}(-)\right]\prod_{l<k}\frac{1}{h(\hat{p}_{l},-\hat{p}_{k})}.
\end{align}
From this, we shall show that
\begin{align}
G(p_{i})\equiv\sum_{{\sf P}_{+}\cup {\sf P_-}=\{1,\ldots,M\}}\left[\prod_{k\in {\sf P}_{-}}(-)\right]\prod_{l<k}\frac{1}{h(\hat{p}_{l},-\hat{p}_{k})}=0
\end{align}
by investigating the poles. The function $G(p_{i})$ has poles at $u_{i}=\pm u_{k}$ because of
\begin{align}
\frac{1}{h(u,v)}=1+\frac{i}{u-v}.
\end{align}
However, these poles are irrelevant since the residues of such poles become zero. Now let us move to the pole at $\hat{p}_{m}=\hat{p}_{n},(n<m)$. Then we can simply show that the residue is
\begin{align}
\sum_{{\sf P}_{+}\cup {\sf P_-}=\{1,\ldots,M\}/\{m,n\}}\left[\prod_{k\in {\sf P}_{-}}(-)\right]&\underset{\hat{p}_{m}\rightarrow \hat{p}_{n}}{\rm res}\Big(\prod_{n\neq i<m}\frac{1}{h(\hat{p}_{i},\pm \hat{p}_{m})}\prod_{m<j}\frac{1}{h(\pm \hat{p}_{m},\hat{p}_{j})}\frac{1}{h(\pm \hat{p}_{n},\pm \hat{p}_{m})}\notag\\
&+\prod_{i<n\neq m}\frac{1}{h(\hat{p}_{i},\pm \hat{p}_{n})}\prod_{n<j}\frac{1}{h(\pm \hat{p}_{n},\hat{p}_{j})}\frac{1}{h(\pm \hat{p}_{m},\pm \hat{p}_{n})}\Big)\notag\\
&=0.
\end{align}
Therefore, the function $G(p_{i})$ does not have any poles. 
Thus, the remaining one is determined by the $u\rightarrow \infty$ behavior. 
Since it trivially becomes zero such as
\begin{align}
G(u_{i}\rightarrow\infty)=\underbrace{1-1+1-1+\cdots}_{2M\ {\rm terms}}=0,
\end{align}
we showed that the function $G(p_{i})$ is precisely zero. This fact means that the summation of the spin-chain length dependent terms doesn't contribute to the structure constants.

As a result, we can written down the final result for the multi-magnon structure constants at tree-level as 
\begin{align}
C_{123}^{M\circ\circ}\propto\prod_{i=1}^{M}{\cal M}(p_{i})e^{i(p_{1}+\cdots+p_{M})\ell_{12}}\sum_{{\sf P}_{+}\cup {\sf P_-}=\{1,\ldots,M\}}\left[\prod_{k\in {\sf P}_{-}}(-e^{2i p_k \ell_{13}})\prod_{l<k}S(p_k,p_l)S(-p_k,p_l)\right]\prod_{i<j}h^{\rm tree}_{YY}(\hat{p}_i,\hat{p}_j)
\end{align}
Furthermore, by dividing the correct norm discussed in section \ref{appg}, we can get the exact form of the structure constants at tree-level including the norm.
We notice that it is nicely matched with our conjecture (\ref{normmulti}).

\subsubsection{Two nontrivial operators with one-magnon : $C^{11\circ}_{123}$ \label{sec324}}

\begin{figure}[t]
\begin{center}
\includegraphics[width=13cm]{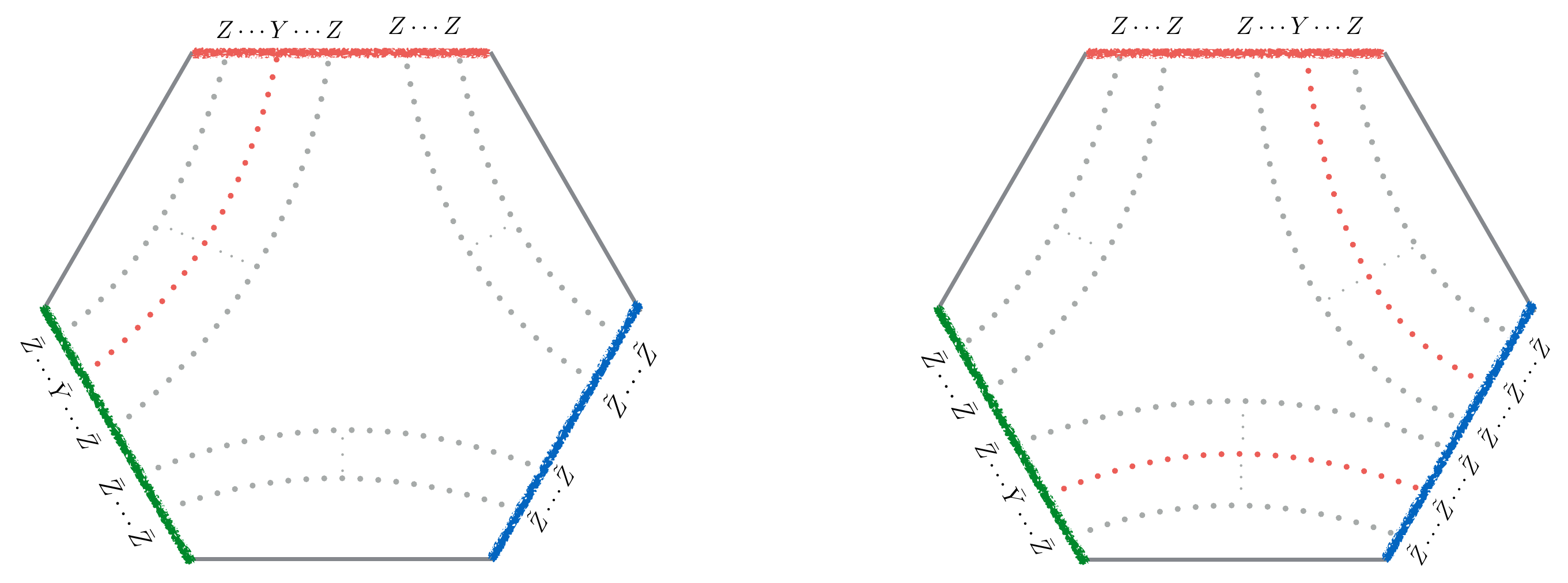}
\caption{$C_{123}^{\rm direct}$ and $C_{123}^{\rm indirect}$}
\label{fig:11c}
\end{center}
\end{figure}

\begin{figure}[t]
\begin{center}
\includegraphics[width=13cm]{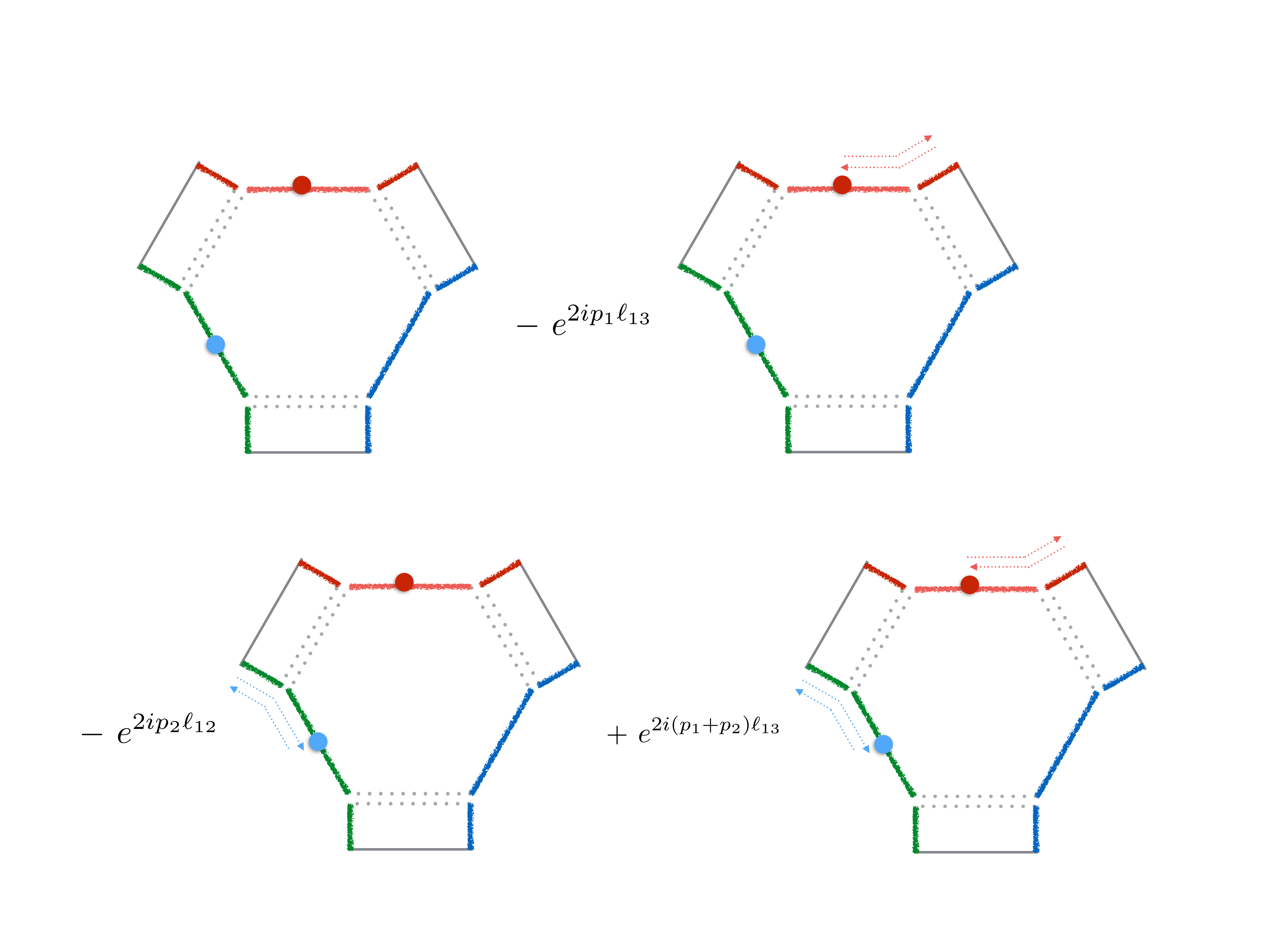}
\caption{By appropriate propagation factors $e^{2ip_{1}\ell_{13}}$ and $e^{2ip_{2}\ell_{12}}$, the each terms can be interpreted as the each magnon with $(p_{1},p_{2}), (-p_{1},p_{2}), (p_{1},-p_{2})$ and $(-p_{1},-p_{2})$ lived on the hexagon.}
\label{fig:p11c}
\end{center}
\end{figure}

We perform computation of tree-level structure constant with an another setup where we consider a situation made of a vacuum state and two nontrivial operators with a magnon respectively. 
For this purpose, we set the first spin-chain to ${\cal O}_{1}:\sum_{x}Z^{x}YZ^{L_{1}-(x+1)}$, the second spin-chain to ${\cal O}_{2}:\sum_{x}\bar{Z}^{x}\bar{Y}\bar{Z}^{L_{2}-(x+1)}$ and the third spin-chain to the vacuum state. With this configuration, there exist two possible ways to contract local operators together for obtaining the structure constant :
\begin{align}
C_{123}^{11\circ}&=C_{123}^{\rm direct}+C_{123}^{\rm indirect},\notag\\
C_{123}^{\rm direct}&\propto\sum_{x=1}^{\ell_{12}}\psi^{(1)}(x,p_{1})\psi^{(1)}(L_{2}-x+1,p_{2}),\notag\\
C_{123}^{\rm indirect}&\propto\sum_{x=\ell_{12}+1}^{L_{1}}\psi^{(1)}(x,p_{1})\sum_{y=1}^{\ell_{23}}\psi^{(1)}(y,p_{2}).
\end{align}
The direct and indirect parts are represented in figure \ref{fig:11c}.
The summation of the propagators in $C_{123}^{\rm direct}$ and in $C_{123}^{\rm indirect}$ are respectively calculated as
\begin{align}
\sum_{x=1}^{\ell_{12}}{\cal A}(x,p_{1}){\cal A}(L_{2}-x+1,p_{2})&={\cal M}(p_{1}){\cal M}(p_{2})\frac{i}{u-v}(e^{ip_{2}L_{2}}-e^{ip_{1}\ell_{12}}e^{ip_{2}\ell_{23}}),\\
-\sum_{x=\ell_{12}+1}^{L_{1}}{\cal A}(x,p_{1})\sum_{y=1}^{\ell_{23}}{\cal A}(y,p_{2})&=-{\cal M}(p_{1}){\cal M}(p_{2})(e^{ip_{1}\ell_{12}}-e^{ip_{1}L_{1}})(1-e^{ip_{2}\ell_{23}}).
\end{align}
By adding the two parts, we have
\begin{align}
C_{123}^{11\circ}&\propto{\cal M}_{\ell_{12}}(p_{1}){\cal M}_{\ell_{23}}(p_{2})\Big[\frac{u-v-i}{u-v}-e^{2ip_{1}\ell_{13}}\frac{-u-v-i}{-u-v}\notag\\
&-e^{2ip_{2}\ell_{12}}\frac{u+v-i}{u+v}+e^{2ip_{1}\ell_{13}}e^{2ip_{2}\ell_{12}}\frac{-u+v-i}{-u+v}\Big], \label{11n}
\end{align}
Even here, we obtained the non-trivial factor in the similar with $C_{123}^{2\circ\circ}$ case, i.e. $\frac{u-v-i}{u-v}$. This factor can be also written by the tree-level hexagon form factor because 
the hexagon form factor $h_{Y|Y}(v|u)$ at the weak coupling regime is simply given by \begin{eqnarray}
h_{Y|Y}(v|u)= \frac{u-v-i}{u-v}+{\cal O}(g).
\end{eqnarray}
Note that one can get $h_{Y|Y}(v|u)$ from the fundamental hexagon form factor $h_{YY}(u,v)$ where all excitations are located at the same physical edges of the hexagon by doing mirror transformations twice. Furthermore the structure constants can also be explained by the dynamical process on the hexagon form factor in figure \ref{fig:p11c}.

\section{Discussion \label{discuss}}

In this paper we gave a conjecture that asymptotic open string three-point function can be decomposed into the one hexagon form factor with boundaries on the mirror edges and we checked its validity by calculating structure constants of local operators inserted on the Wilson loop at tree-level.
The local operators are interpreted as open spin-chain with integrable boundary conditions. Based on this, we could construct the Bethe wave functions, and compute contractions between local operators. 

Since our check was restricted to the weak coupling regime, it would be very urgent to understand if our conjecture is indeed working in integrable bootstrap. The most convincing direction is to reproduce two-loop results in \cite{KKKN} through the hexagon approach. 
Because the setup of \cite{KKKN} has no any local operator insertion which means absence of physical magnon, 
we might directly detect mirror particle contributions and properties of boundaries.
Furthermore, there would be other interesting directions to develop. Let us introduce some possible future works.\\

{\bf Higher-rank sectors}\\

As a nontrivial check for this paper, it would be interesting to study on the higher-rank sectors where we need to consider the nested technique. 
Beyond the rank-one sector, we generally have to consider the same kind of scalar fields coupled to the Wilson loops as the excitations on one of three open chains. Therefore, their boundary conditions become highly nontrivial since the reflection matrix has generally a nondiagonal form. 
The quantum exact reflection matrix is already available from \cite{Drukker:2012de, Correa:2012hh}. Therefore, if we diagonalize the double row transfer matrix with the reflection matrix and obtain the Bethe wave functions, one would get the tree-level structure constants for higher-rank sectors. 
Then it would be very important to understand if any new objects would contribute in bootstrap, and how to glue the boundaries in general. \\

{\bf Other integrable open strings}\\

There are many other integrable open string configurations in $AdS/CFT$. They usually attach to specific $D$-branes, which preserve the bulk integrability even though the bulk symmetry group is reduced to a smaller group. Such integrable open strings have corresponding dual gauge invariant operators. 
One of them is the integrable open string attached to the maximal giant graviton in planar limit. Its dual gauge invariant operator is given by determinant type operator.\footnote{Giant graviton has ${\cal O}(N)$ $R$-charge by definition. If the number of giant gravitons scales as ${\cal O}(N)$ in the large $N$ limit, the original $AdS_{5} \times S^{5}$ background is deformed by gravity backreaction. Such a new geometry is described by LLM geometry \cite{llm}. Then, the dual gauge invariant operators are represented by the Schur polynomials or the restricted Schur polynomials \cite{geoy}. Even though integrability in the LLM geometry is generally broken, we can still have some nontrivial integrable cases \cite{llmmagnon}.} 
As the open spin-chain Hamiltonian is also known for this operator, the tree level analysis done in this paper could be performed similarly.  
For example, since open strings attached to the $Y=0$ brane have a diagonal reflection matrix, the higher-rank sector could be studied simpler than the Wilson loops with operator insertions. In addition, as the $D$-brane is explicitly given in this case, the existence of new objects in bootstrap may be directly uncovered.\footnote{On the other hand, $D$-brane was hidden in the $AdS$ boundary for the open strings dual to the Wilson loops.}  \\

{\bf Application to other theories}\\

It may be useful to apply the hexagon decomposition method to other theories. For example, as a simple integrable $CFT$ model, there is the three-dimensional $O(N)$ vector model as free boson (fermion) theory. The theory is trivially integrable since it is a free theory. 
The simplest dual gauge invariant operator is given by the bi-fundamental operator. 
The operator would map to a sort of open spin-chain with only two sites. Furthermore, this theory has an attractive property, so-called $3d$-bosonization \cite{Maldacena:2011jn, Maldacena:2012sf}. The boson theory and a fermion theory can be interchanged by the Cherm-Simons coupling constants. 
How this information is realized in the hexagon framework would be an interesting question. 
The $AdS$ dual description of this model is the Vasiliev's higher spin theory which is defined as a bulk theory \cite{Klebanov:2002ja, Giombi:2009wh, Giombi:2010vg, Giombi:2012ms}. The coupling to
Chern-Simons amounts to changing the boundary condition on the Vasiliev
side. On the other hand, the hexagon method is based on the world-sheet formalism, since the hexagon is basically obtained by the cutting the world-sheet. Therefore, if we can express the three-point function of the three dimensional $O(N)$ vector model in terms of the hexagon method, it may give a useful hint for the world-sheet formulation of the higher spin theory.

\subsection*{Acknowledgement}

We thank Shota Komatsu for helpful discussions. 
We also thank Zoltan Bajnok, Robert de Mello Koch, Rafael Nepomechie and Hendrik J.R. van Zyl for valuable comments.
N.K. gratefully acknowledges T. Nishimura for useful conversations. 
The research of M.K. was supported by a postdoctoral fellowship of the Hungarian Academy of Sciences, a Lend$\ddot{\rm u}$let grant, OTKA 116505 and by the South African Research Chairs Initiative of the Department of Science and Technology and National Research
Foundation, whereas the researches of N.K. is supported in part by JSPS Research Fellowship for Young Scientists, from the Japan Ministry of Education, Culture, Sports, Science and Technology.

\appendix

\section{Properties of the wave function of the open spin-chain \label{appa}}
Here we summarize some important properties of the wave function of the open spin-chain appearing in our setup.

\subsection{One-magnon}

By acting with the Hamiltonian on the wave function, the basis of the wave function would be mixed. 
The L.H.S. of (\ref{eigen1}) becomes a linear combination of the basis elements. Thereby, we obtain the following conditions for the wave function :
\begin{align}
&E^{(1)}\psi(1)=(1+C_{1})\psi(1)-\psi(2),\\
&E^{(1)}\psi(x)=2\psi(x)-\psi(x-1)-\psi(x+1),\\
&E^{(1)}\psi(L)=(1+C_{L})\psi(L)-\psi(L-1).
\end{align}
By solving these equations, we can get the boundary conditions as below.
\begin{align}
\psi(0)=(1-C_{1})\psi(1),\ \ \psi(L+1)=(1-C_{L})\psi(L)
\end{align}
Furthermore, as we insert the boundary conditions to the wave function (\ref{1wave}),  we obtain the Bethe equation :
\begin{align}
1=e^{2ipL}B_{1}(-p)B_{L}(p)
\end{align}
where $B_{1}(p)$ and $B_{L}(p)$ are defined as
\begin{align}
B_{1}(p)&\equiv -e^{-ip}\frac{A(-p)}{A(p)}=-\frac{e^{-ip}-(1-C_{1})}{1-(1-C_{1})e^{-ip}}\notag,\\
B_{L}(p)&\equiv  -e^{-ip}e^{-2ipL}\frac{A(-p)}{A(p)}=-\frac{e^{ip}-(1-C_{L})}{1-(1-C_{L})e^{ip}}.
\end{align}

\subsection{Two-magnon}

The Bethe wave function for two-magnon satisfies the following conditions :
\begin{align}
E^{(2)}\psi(1,x_{2})&=(1+C_{1})\psi(1,x_{2})+2\psi(1,x_{2})-\psi(2,x_{2})\notag\\
&-\psi(1,x_{2}-1)-\psi(1,x_{2}+1),\\
E^{(2)}\psi(x_{1},x_{2})&=4\psi(x_{1},x_{2})-\psi(x_{1}-1,x_{2})-\psi(x_{1}+1,x_{2})\notag \\
&-\psi(x_{1},x_{2}-1)-\psi(x_{1},x_{2}+1),\label{b1}\\
E^{(2)}\psi(x_{1},x_{1}+1)&=2\psi(x_{1},x_{1}+1)-\psi(x_{1}-1,x_{1}+1)-\psi(x_{1},x_{2}+2),\label{b2}\\
E^{(2)}\psi(x_{1},L)&=(1+C_{L})\psi(x_{1},L)+2\psi(x_{1},L)-\psi(x_{1},L-1)\notag\\
&-\psi(x_{1}+1,L_{1})-\psi(x_{1}-1,L_{1}).
\end{align}
From these equations, we obtain the boundary conditions as follows. 
\begin{align}
\psi(0,x_{2})=(1-C_{1})\psi(1,x_{2}),\ \ \psi(x_{1},L+1)=(1-C_{L})\psi(x_{1},L)
\end{align}
The constraints (\ref{b1}) and (\ref{b2}) for the bulk wave functions give the following equation :
\begin{align}
0=2\psi(x_{1},x_{1}+1)-\psi(x_{1}+1,x_{1}+1)-\psi(x_{1},x_{1}).
\end{align}
By solving the equation, we obtain the bulk $S$-matrix 
\begin{align}
S(p_{2},p_{1})&=\frac{u-v-i}{u-v+i},
\end{align}
and boundary coefficients.
Finally, the Bethe equation for two-magnon is given as
\begin{align}
1=e^{-2iLp_{2}}B_{1}(p_{2})B_{L}(-p_{2})S(p_{2},p_{1})S(p_{2},-p_{1}).
\end{align}

Next let us comment on the difference between the notations of \cite{OTY} and our notations. In \cite{OTY}, the wave function was written as 
\begin{align*}
\psi&=A(p_{1},p_{2})e^{i(p_{1}x_{1}+p_{2}x_{2})}-A(-p_{1},p_{2})e^{-i(p_{1}x_{1}-p_{2}x_{2})}-A(p_{1},-p_{2})e^{i(p_{1}x_{1}-p_{2}x_{2})}\\
&+A(-p_{1},-p_{2})e^{-i(p_{1}x_{1}+p_{2}x_{2})}-A(p_{2},p_{1})e^{i(p_{2}x_{1}+p_{1}x_{2})}+A(p_{2},-p_{1})e^{-i(p_{2}x_{1}-p_{1}x_{2})}\\
&+A(-p_{2},p_{1})e^{i(p_{2}x_{1}-p_{1}x_{2})}-A(-p_{2},-p_{1})e^{-i(p_{2}x_{1}+p_{1}x_{2})}.
\end{align*}
By considering the half-step shift,\footnote{we identify boundaries of spin-chain with the half-step shifted points from the first and the last sites (i.e., the‘ $1/2$-th ’and the‘$L+1/2$-th ’sites).} we have
\begin{align*}
\psi^{(2)}&=A^{'}(p_{1},p_{2})e^{i\left(p_{1}(x_{1}-\frac{1}{2})+p_{2}(x_{2}-\frac{1}{2})\right)}-A^{'}(-p_{1},p_{2})e^{-i\left(p_{1}(x_{1}-\frac{1}{2})-p_{2}(x_{2}-\frac{1}{2})\right)}\\
&-A^{'}(p_{1},-p_{2})e^{i\left(p_{1}(x_{1}-\frac{1}{2})-p_{2}(x_{2}-\frac{1}{2})\right)}+A^{'}(-p_{1},-p_{2})e^{-i\left(p_{1}(x_{1}-\frac{1}{2})+p_{2}(x_{2}-\frac{1}{2})\right)}\\
&-A^{'}(p_{2},p_{1})e^{i\left(p_{2}(x-\frac{1}{2})_{1}+p_{1}(x_{2}-\frac{1}{2})\right)}+A^{'}(p_{2},-p_{1})e^{i\left(p_{2}(x_{1}-\frac{1}{2})-p_{1}(x_{2}-\frac{1}{2})\right)}\\
&+A^{'}(-p_{2},p_{1})e^{-i\left(p_{2}(x_{1}-\frac{1}{2})-p_{1}(x_{2}-\frac{1}{2})\right)}-A^{'}(-p_{2},-p_{1})e^{-i\left(p_{2}(x_{1}-\frac{1}{2})+p_{1}(x_{2}-\frac{1}{2})\right)}\label{wave}
\end{align*}
where 
\begin{align*}
A^{'}(p_{1},p_{2})=e^{\frac{i}{2}p_{1}+\frac{i}{2}p_{2}}A(p_{1},p_{2}).
\end{align*}
By using the Bethe equation, bulk and boundary $S$-matrices $S(p_{j},p_{i})=\frac{A(p_{j},p_{i})}{A(p_{i},p_{j})}$, $B_{L}(p_{1})=-e^{-ip_{1}}e^{-2ip_{1}L}\frac{A(p_{2},-p_{1})}{A(p_{2},p_{1})}$ and $B_{L}(p_{2})=-e^{-ip_{2}}e^{-2ip_{2}L}\frac{A(p_{1},-p_{2})}{A(p_{1},p_{2})}$ \footnote{$B_{1}(p_{1})=-e^{-ip_{1}}\frac{A(-p_{1},p_{2})}{A(p_{1},p_{2})}$ and $B_{1}(p_{2})=-e^{-ip_{2}}\frac{A(-p_{2},p_{1})}{A(p_{2},p_{1})}$}, we get our notation
\begin{align*}
&\psi^{(2)}(x_{1},x_{2})/A'(p_{1},p_{2})=\notag\\
&g(x_{1},p_{1};x_{2},p_{2})+S(p_{2},p_{1})S(-p_{2},p_{1})e^{2ip_{1}L}B_{L}(p_{1})g(x_{1},-p_{1};x_{2},p_{2})\notag\\
&+e^{2ip_{2}L}B_{L}(p_{2})g(x_{1},p_{1};x_{2},-p_{2})+S(p_{2},p_{1})S(-p_{2},p_{1})e^{2i(p_{1}+p_{2})L}B_{L}(p_{1})B_{L}(p_{2})g(x_{1},-p_{1};x_{2},-p_{2})
\end{align*}
where the function $g(x_{1},p_{1};x_{2},p_{2})$ is defined as
\begin{align*}
g(x_{1},p_{1};x_{2},p_{2})&\equiv{\cal A}(x_{1},p_{1}){\cal A}(x_{2},p_{2})+S(p_{2},p_{1}){\cal A}(x_{1},p_{2}){\cal A}(x_{2},p_{1}).
\end{align*}

\section{Hexagon form factor as the weight factor \label{appb}}

\begin{figure}[t]
\begin{center}
\includegraphics[width=10cm]{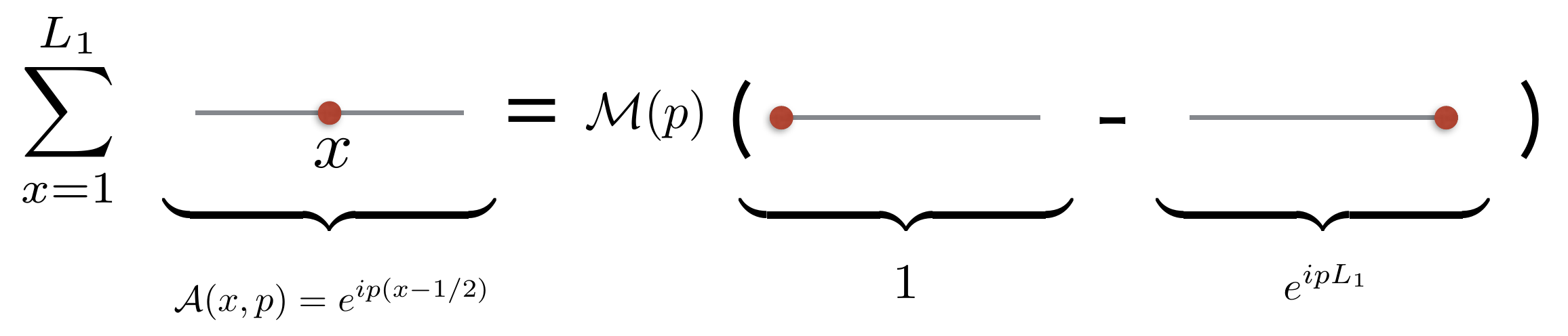}
\caption{Summation over the position as dynamical processes of magnons.}
\label{fig:we}
\end{center}
\end{figure}

The structure constants can be expressed in terms of contributions obtained by geometric series. The geometric series is deeply related to dynamical processes of magnons, namely how their propagations of magnons are given before and after summation. 
The fundamental contribution is given as 
\begin{align}
\sum_{x=1}^{L_{1}}{\cal A}(x,p)={\cal M}(p)(1-e^{ipL_{1}}).
\end{align}
The summation over the position that the magnons are located can be described as in figure \ref{fig:we}.

Based on this summation rule, the expression for two-magnon becomes
\begin{align}
\sum_{1\leq x_{1}<x_{2}\leq L_{1}}{\cal A}(x_{1},p_{1}){\cal A}(x_{2},p_{2})&=\sum_{x_{1}=1}^{L_{1}}\sum_{x_{2}=x_{1}+1}^{L_{1}}{\cal A}(x_{1},p_{1}){\cal A}(x_{2},p_{2})\notag\\
&={\cal M}(p_{1}){\cal M}(p_{2}) \Big\{\frac{i+2v}{2(u+v)}(1-e^{i(p_{1}+p_{2})L_{1}})-(e^{ip_{2}L_{1}}-e^{i(p_{1}+p_{2})L_{1}})\Big\}, \label{sumAA}
\end{align}
where the propagation factor such as $e^{i p_{j} L}$ is included if the magnon moves on the distance $L$. On the other hand, as the cluster of the magnons more than one magnon can move, it will get a weight factor.

\begin{figure}[t]
\begin{center}
\includegraphics[width=13cm]{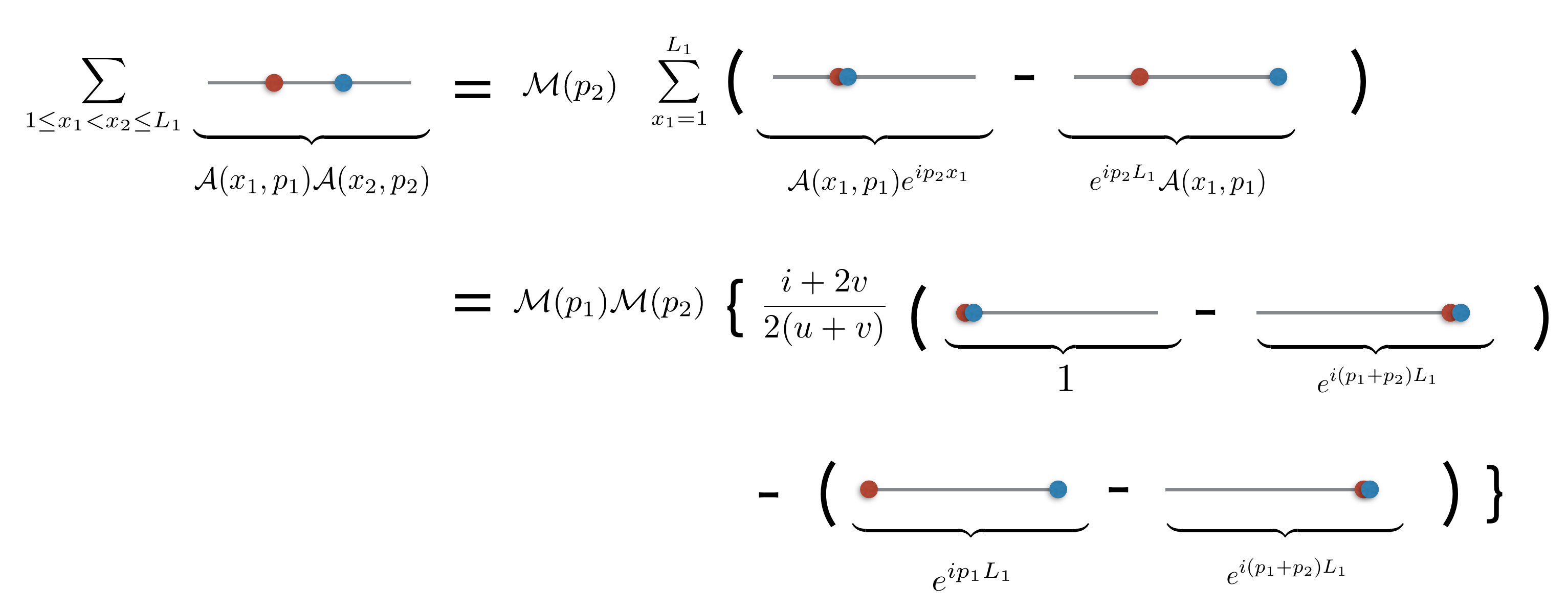}
\caption{Weight factor in case of the two-magnon}
\label{fig:we2}
\end{center}
\end{figure}

Let us explain in detail. 
At the start, the magnons with momenta $p_{1,2}$ be located at their positions $x_{1,2}$ and we try to compute the RHS in (\ref{sumAA}). First, by summation for the position $x_{2}$,
\begin{align}
\sum_{x_{2}=x_{1}+1}^{L_{1}}{\cal A}(x_{2},p_{2})={\cal M}(p_{2})(e^{ip_{2}x_{1}}-e^{ip_{2}L_{1}}),
\end{align}
the magnon with momentum $p_{2}$ can move to the position $\ell_{12}$ and $L_{1}$, see first line in figure \ref{fig:we2}. Next, the summation for the position $x_{1}$ also include the propagation factor:
\begin{align}
{\cal M}(p_{2})\sum_{x_{1}=1}^{L_{1}}{\cal A}(x_{1},p_{1})e^{ip_{2}x_{1}}&={\cal M}(p_{1}){\cal M}(p_{2})\frac{i+2v}{2(u+v)}(1-e^{i(p_{1}+p_{2})L_{1}}),\\
{\cal M}(p_{2})e^{ip_{2}L_{1}}\sum_{x_{1}=1}^{L_{1}}{\cal A}(x_{1},p_{1})&={\cal M}(p_{1}){\cal M}(p_{2})(e^{ip_{2}L_{1}}-e^{i(p_{1}+p_{2})L_{1}}).
\end{align}
Then, since two magnons move at the same time, the non-trivial weight factor is raised, where the factor for the two-magnon case is given as $\frac{i+2v}{2(u+v)}$ as described in figure \ref{fig:we2}.
Combining the weight factor with the $S$-matrix, we can get the tree-level hexagon form factor given as
\begin{align}
\frac{i+2v}{2(u+v)}+S(v,u)\frac{i+2u}{2(u+v)}=h_{YY}^{\rm tree}(u,v).
\end{align}
Thus, the hexagon form factor can be understood as the weight factors for the dynamics of magnons.

\begin{figure}[t]
\begin{center}
\includegraphics[width=14cm]{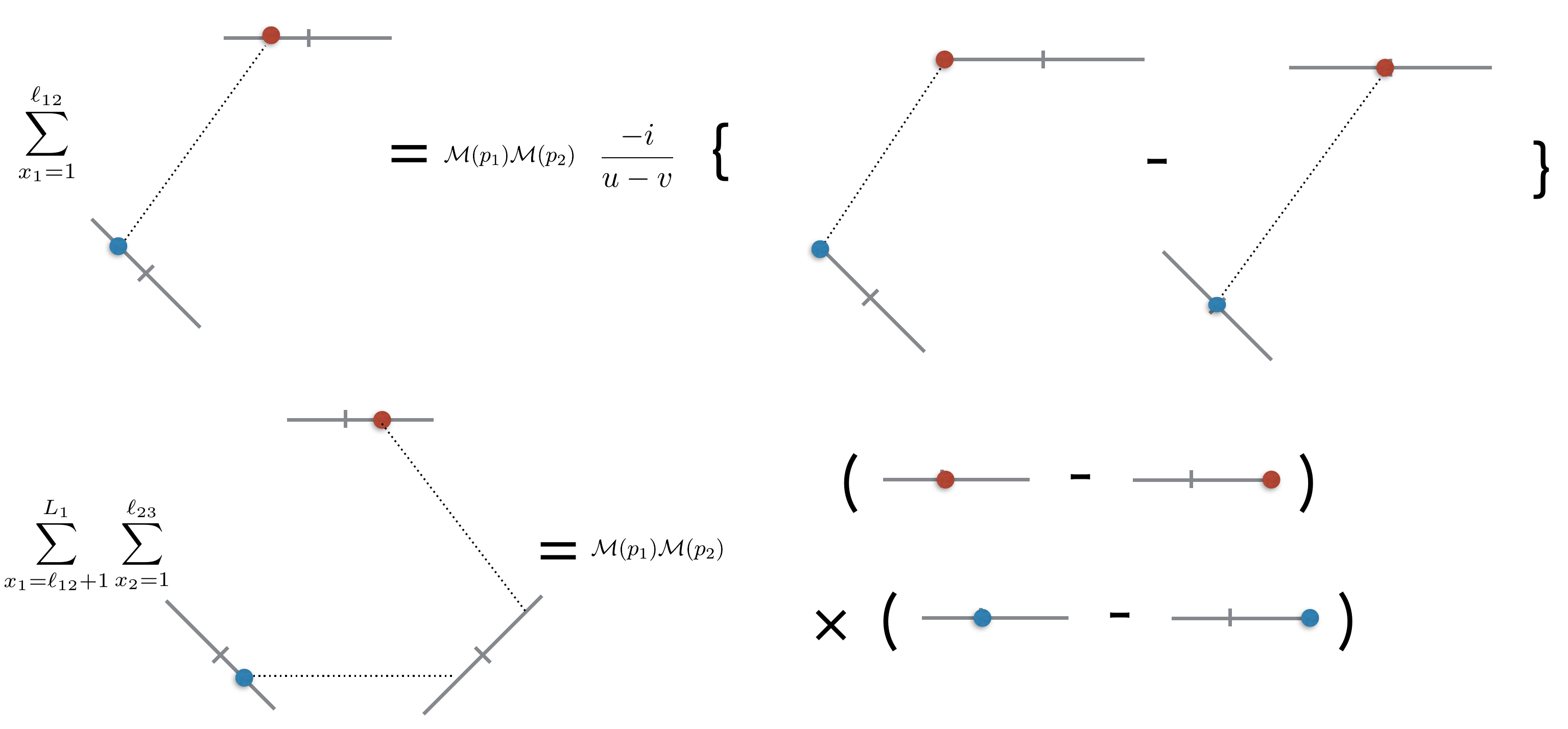}
\caption{The direct and the indirect contributions}
\label{fig:we11}
\end{center}
\end{figure}

Now let us consider the case of two operators with a magnon separately. 
Here the contribution is divided into two parts : direct part and indirect part. 
The indirect part can be written as
\begin{align}
-\sum_{x_{1}=\ell_{12}+1}^{L_{1}}\sum_{x_{1}=1}^{\ell_{23}}{\cal A}(x_{1},p_{1}){\cal A}(x_{2},p_{2})=-{\cal M}(p_{1}){\cal M}(p_{2})(e^{ip_{1}\ell_{12}}-e^{ip_{1}L_{1}})(1-e^{ip_{2}\ell_{23}}).
\end{align}
In this case, non-trivial factor doesn't appear since the magnons move independently. 
On the other hand, the direct part becomes
\begin{align}
\sum_{x_{1}=1}^{\ell_{12}}{\cal A}(x_{1},p_{1}){\cal A}(L_{2}-x_{1}+1,p_{2})={\cal M}(p_{1}){\cal M}(p_{2})\frac{i}{u-v}(e^{ip_{1}L_{2}}-e^{ip_{1}\ell_{12}}e^{ip_{2}\ell_{23}}).
\end{align}
The direct and indirect parts are pictorially described in figure \ref{fig:we11}.
Furthermore, by adding the weight factors, we get the hexagon form factor
\begin{align}
1-\frac{i}{u-v}=h^{\rm tree}_{Y|Y}(v|u).
\end{align}

\section{$M$-particle hexagon form factor \label{appc}}

In section \ref{sec323}, we have discussed about the structure constants of $M$-magnon. 
However, as the general configuration is quite involved,  
let us first consider the case of a few number of magnons. 
For one-magnon, the part of the structure constants is written as
\begin{align}
\sum^{L_{1}}_{x_{1}=\ell_{12}+1}{\cal A}(x,p)&=e^{-\frac{i}{2}p}\frac{1}{e^{-ip}-1}e^{ip\ell_{12}}+\cdots\notag\\
&={\cal M}(p)h(u)e^{ip\ell_{12}}+\cdots .
\end{align}
We can see that we get the one-particle hexagon form factor. Next, for two-magnon, we have
\begin{align}
&\sum^{L_{1}}_{x_{1}=\ell_{12}+1}\sum_{x_{2}=x_{1}+1}^{L_{1}}\Big({\cal A}(x_{1},p_{1}){\cal A}(x_{2},p_{2})+S(p_{2},p_{1}){\cal A}(x_{1},p_{2}){\cal A}(x_{2},p_{1})\Big)\notag\\
&=e^{-\frac{i}{2}(p_{1}+p_{2})}h(p_{1},p_{2})\notag\\
&\Big(\frac{1}{h(p_{1},p_{2})}\frac{1}{e^{-ip_{2}}-1}\frac{1}{e^{-i(p_{2}+p_{1})}-1}+\frac{1}{h(p_{2},p_{1})}\frac{1}{e^{-ip_{1}}-1}\frac{1}{e^{-i(p_{2}+p_{1})}-1}\Big)e^{i(p_{1}+p_{2})\ell_{12}}+\cdots\notag\\
&={\cal M}(p_{1}){\cal M}(p_{2})h(u_{1},u_{2})e^{i(p_{1}+p_{2})\ell_{12}}+\cdots ,
\end{align}
where in the second line we used the property $S(v,u)=\frac{h(u,v)}{h(v,u)}$ and the terms in bracket become $\frac{1}{(e^{-ip_{1}}-1)(e^{-ip_{2}}-1)}$. 
Next, the three-particle hexagon form factor can be written as
\begin{align}
&\sum^{L_{1}}_{x_{1}=\ell_{12}+1}\sum_{x_{2}=x_{1}+1}^{L_{1}}\sum_{x_{3}=x_{2}+1}^{L_{1}}\Big({\cal A}(x_{1},p_{1}){\cal A}(x_{2},p_{2}){\cal A}(x_{3},p_{3})+S(p_{2},p_{1}){\cal A}(x_{1},p_{2}){\cal A}(x_{2},p_{1}){\cal A}(x_{3},p_{3})\Big)\notag\\
&+S(p_{3},p_{2}){\cal A}(x_{1},p_{1}){\cal A}(x_{2},p_{3}){\cal A}(x_{3},p_{2})+S(p_{3},p_{1}){\cal A}(x_{1},p_{3}){\cal A}(x_{2},p_{2}){\cal A}(x_{3},p_{1})\notag\\
&+S(p_{2},p_{1})S(p_{3},p_{1}){\cal A}(x_{1},p_{2}){\cal A}(x_{2},p_{3}){\cal A}(x_{3},p_{1})+S(p_{3},p_{2})S(p_{3},p_{1}){\cal A}(x_{1},p_{3}){\cal A}(x_{2},p_{1}){\cal A}(x_{3},p_{2}),
\end{align}
where we used the following property of the summation
\begin{align}
&\sum^{L_{1}}_{x_{1}=\ell_{12}+1}\sum_{x_{2}=x_{1}+1}^{L_{1}}\sum_{x_{3}=x_{2}+1}^{L_{1}}{\cal A}(x_{1},p_{1}){\cal A}(x_{2},p_{2}){\cal A}(x_{3},p_{3})\notag\\
&=e^{-\frac{i}{2}(p_{1}+p_{2}+p_{3})}\frac{1}{e^{-ip_{3}}-1}\frac{1}{e^{-i(p_{3}+p_{2})}-1}\frac{1}{e^{-i(p_{3}+p_{2}+p_{1})}-1}e^{i(p_{1}+p_{2}+p_{3})\ell_{12}}+\cdots .
\end{align}
For the bridge length dependent terms, we have
\begin{align}
&e^{-\frac{i}{2}(p_{1}+p_{2}+p_{3})}h(p_{1},p_{2})h(p_{1},p_{3})h(p_{2},p_{3})\notag\\
&\Big\{\frac{{\cal M}(p_{1},p_{2},p_{3})}{h(p_{1},p_{2})h(p_{2},p_{3})h(p_{1},p_{3})}+\frac{{\cal M}(p_{2},p_{1},p_{3})}{h(p_{2},p_{1})h(p_{1},p_{3})h(p_{2},p_{3})}+\frac{{\cal M}(p_{1},p_{3},p_{2})}{h(p_{1},p_{2})h(p_{1},p_{3})h(p_{3},p_{2})}\notag\\
&+\frac{{\cal M}(p_{3},p_{2},p_{1})}{h(p_{1},p_{2})h(p_{3},p_{1})h(p_{2},p_{3})}+\frac{{\cal M}(p_{2},p_{3},p_{1})}{h(p_{2},p_{1})h(p_{3},p_{1})h(p_{2},p_{3})}+\frac{{\cal M}(p_{3},p_{1},p_{2})}{h(p_{1},p_{2})h(p_{3},p_{2})h(p_{3},p_{1})}\Big\}.
\end{align}
Using the identity
\begin{align}
M(p_{1},p_{2},p_{3})=\frac{{\cal M}(p_{2},p_{3})}{e^{-i(p_{1}+p_{2}+p_{3})}-1},
\end{align}
we get
\begin{align}
&e^{-\frac{i}{2}(p_{1}+p_{2}+p_{3})}h(p_{1},p_{2})h(p_{1},p_{3})h(p_{2},p_{3})\frac{1}{e^{-i(p_{1}+p_{2}+p_{3})}}\notag\\
&\Big\{\left(\frac{{\cal M}(p_{2},p_{3})}{h(p_{2},p_{3})}+\frac{{\cal M}(p_{3},p_{2})}{h(p_{3},p_{2})}\right)\frac{1}{h(p_{1},p_{2})h(p_{1},p_{3})}+\left(\frac{{\cal M}(p_{1},p_{2})}{h(p_{1},p_{2})}+\frac{{\cal M}(p_{2},p_{1})}{h(p_{2},p_{1})}\right)\frac{1}{h(p_{3},p_{1})h(p_{3},p_{2})}\notag\\
&+\left(\frac{{\cal M}(p_{1},p_{3})}{h(p_{1},p_{3})}+\frac{{\cal M}(p_{3},p_{1})}{h(p_{3},p_{1})}\right)\frac{1}{h(p_{2},p_{3})h(p_{2},p_{1})}
\Big\}.
\end{align}
By mathematical induction, we have
\begin{align}
&e^{-\frac{i}{2}(p_{1}+p_{2}+p_{3})}h(p_{1},p_{2})h(p_{1},p_{3})h(p_{2},p_{3})\frac{1}{e^{-i(p_{1}+p_{2}+p_{3})}}\frac{1}{(e^{-ip_{1}}-1)(e^{-ip_{2}}-1)(e^{-ip_{3}}-1)}\notag\\
&\Big\{\frac{e^{-ip_{1}}-1}{h(p_{1},p_{2})h(p_{1},p_{3})}+\frac{e^{-ip_{3}}-1}{h(p_{3},p_{1})h(p_{3},p_{2})}+\frac{e^{-ip_{2}}-1}{h(p_{2},p_{3})h(p_{2},p_{1})}
\Big\}.
\end{align}
Finally, by using the following residue integral
\begin{align}
\oint \frac{dz}{2\pi i}\frac{1}{z}\left(\prod_{k=1}^{3}\frac{u_{k}-z-i/2}{u_{k}-z+i/2}-1\right),
\end{align}
we find the relation
\begin{align}
\frac{1}{e^{-i(p_{1}+p_{2}+p_{3})}}=\frac{e^{-ip_{1}}-1}{h(p_{1},p_{2})h(p_{1},p_{3})}+\frac{e^{-ip_{3}}-1}{h(p_{3},p_{1})h(p_{3},p_{2})}+\frac{e^{-ip_{2}}-1}{h(p_{2},p_{3})h(p_{2},p_{1})}.
\end{align}
Therefore we get the following expression for three-particle hexagon form factor :
\begin{align}
{\cal M}(p_{1}){\cal M}(p_{2}){\cal M}(p_{3})h(u_{1},u_{2},u_{3}).
\end{align}
As a result, we expect that the $M$-particle hexagon form factor can be expressed as
\begin{align}
&h_{YY}(u_{1},\cdots,u_{M})\notag\\
&=\frac{1}{{\cal M}(p_{1})\cdots {\cal M}(p_{1})}\sum_{x_{1}<\cdots<x_{M}}\sum_{\sigma_{1}\neq\cdots\neq\sigma_{M}}\prod_{\underset{j<k}{\sigma_{k}<\sigma_{j}}}S(p_{\sigma_{k}},p_{\sigma_{k}})\prod_{l=1}^{M}{\cal A}(x_{l},p_{\sigma_{l}}).
\end{align}

\section{Bridge length independent terms \label{appd}}

In this appendix, we show that the bridge length independent terms (\ref{no}) are cancelled by each other for a few number of magnons. 
The key points are as follows: First, we divide the terms by the factor $\prod_{l<k}h(p_{l},p_{k})h(p_{l},-p_{k})$. Second, by investigating the poles, we find that these are spurious poles. Finally, we check that the constant parts at $u\rightarrow\infty$ are exactly cancelled out.

Now let us consider two-magnon case,\footnote{since one-magnon case has no poles, it is trivial.} then the terms are given by
\begin{align}
&\sum\psi^{(2)}(x_{1},x_{2})/{\cal M}(p_{1}){\cal M}(p_{2})e^{i(p_{1}+p_{2})L}|_{\ell_{ij}=0}
\\
&=h(u_{1},u_{2})-S(u_{2},u_{1})S(-u_{2},u_{1})h(-u_{1},u_{2})-h(u_{1},-u_{2})+S(u_{2},u_{1})S(-u_{2},u_{1})h(-u_{1},-u_{2}),\notag
\end{align}
which can be straightforwardly rewritten as
\begin{align}
h(u_{1},u_{2})h(u_{1},-u_{2})\left(\frac{1}{h(u_{1},-u_{2})}-\frac{1}{h(-u_{1},-u_{2})}-\frac{1}{h(u_{1},u_{2})}+\frac{1}{h(-u_{1},u_{2})}\right).
\end{align}
Although it seems that we have poles at $u_{1}=\pm u_{2}$, their residues are all zero:
\begin{align}
\underset{u\rightarrow v}{\rm res}\left(\frac{1}{h(u_{1},\pm u_{2})}+\frac{1}{h(-u_{1},\mp u_{2})}\right)=i-i=0.
\end{align}
Therefore these are spurious poles. Furthermore the $u\rightarrow\infty$ behavior becomes trivially zero. 
As a result, we showed that the spin chain length dependent term is exactly zero for the two-magnon case.

For the three-magnon case, the terms are given by
\begin{align*}
&\sum\psi^{(3)}(x_{1},x_{2},x_{3})/{\cal M}(u_{1}){\cal M}(u_{2}){\cal M}(u_{3}){\cal M}(u_{3})e^{i(p_{1}+p_{2}+p_{3})L}|_{\ell_{ij}=0}\\
&=h(u_{1},u_{2},u_{3})-h(u_{1},u_{2},-u_{3})\\
&-S(u_{2},u_{1})S(-u_{2},u_{1})S(u_{3},u_{1})S(-u_{3},u_{1})\{h(-u_{1},u_{2},u_{3})-h(-u_{1},u_{2},-u_{3})\}\\
&-S(u_{3},u_{2})S(-u_{3},u_{2})\{h(u_{1},-u_{2},u_{3})-h(u_{1},-u_{2},-u_{3})\}\\
&+S(u_{3},u_{2})S(-u_{3},u_{2})S(u_{2},u_{1})S(-u_{2},u_{1})S(u_{3},u_{1})S(-u_{3},u_{1})\{h(-u_{1},-u_{2},u_{3})-h(-u_{1},-u_{2},-u_{3})\}.
\end{align*}
By taking $\prod_{i<j}h(u_{i},u_{j})h(u_{i},-u_{j})$ in front of whole expression,  we could obtain
\begin{align}
&h(u_{2},u_{3})h(u_{2},-u_{3})h(u_{1},u_{2})h(u_{1},-u_{2})h(u_{1},u_{3})h(u_{1},-u_{3})\notag\\
&\Big(\frac{1}{h(u_{1},u_{2})h(u_{1},-u_{3})h(u_{3},u_{2})}-\frac{1}{h(u_{1},u_{2})h(u_{1},u_{3})h(-u_{3},u_{2})}\notag\\
&-\frac{1}{h(-u_{1},u_{2})h(-u_{1},-u_{3})h(u_{3},u_{2})}+\frac{1}{h(-u_{1},u_{2})h(-u_{1},u_{3})h(-u_{3},u_{2})}\notag\\
&-\frac{1}{h(u_{1},-u_{2})h(u_{1},-u_{3})h(u_{3},-u_{2})}+\frac{1}{h(u_{1},-u_{2})h(u_{1},u_{3})h(-u_{3},-u_{2})}\notag\\
&+\frac{1}{h(-u_{1},-u_{2})h(-u_{1},-u_{3})h(u_{3},-u_{2})}-\frac{1}{h(-u_{1},-u_{2})h(-u_{1},u_{3})h(-u_{3},-u_{2})}\Big).
\end{align}
Similarly, this expression appears to have poles at $u_{1}=\pm u_{2},u_{1}=\pm u_{3}$ and $u_{2}=\pm u_{3}$. However, the residues are zero again:
\begin{align*}
\underset{u_{1}\rightarrow \pm u_{2}}{\rm res}&\Big(\frac{1}{h(u_{1},\pm u_{2})h(u_{1},-u_{3})h(u_{3},\pm u_{2})}-\frac{1}{h(u_{1},\pm u_{2})h(u_{1},u_{3})h(-u_{3},\pm u_{2})}\notag\\
&+\frac{1}{h(-u_{1},\mp u_{2})h(-u_{1},-u_{3})h(u_{3},\mp u_{2})}-\frac{1}{h(-u_{1},\mp u_{2})h(-u_{1},u_{3})h(-u_{3},\mp u_{2})}\Big)=0,\notag\\
\underset{u_{1}\rightarrow \pm u_{3}}{\rm res}&\Big(\frac{1}{h(u_{1},u_{2})h(u_{1},\pm u_{3})h(\mp u_{3},u_{2})}-\frac{1}{h(u_{1},-u_{2})h(u_{1},\pm u_{3})h(\mp u_{3},-u_{2})}\notag\\
&+\frac{1}{h(-u_{1},u_{2})h(-u_{1},\mp u_{3})h(\pm u_{3},u_{2})}-\frac{1}{h(-u_{1},-u_{2})h(-u_{1},\mp u_{3})h(\pm u_{3},-u_{2})}\Big)=0,\notag\\
\underset{u_{2}\rightarrow \pm u_{3}}{\rm res}&\Big(\frac{1}{h(u_{1},u_{2})h(u_{1},\mp u_{3})h(\pm u_{3},u_{2})}-\frac{1}{h(-u_{1},u_{2})h(-u_{1},\mp u_{3})h(\pm u_{3},u_{2})}\notag\\
&+\frac{1}{h(u_{1},-u_{2})h(u_{1},\pm u_{3})h(\mp u_{3},-u_{2})}-\frac{1}{h(-u_{1},-u_{2})h(-u_{1},\pm u_{3})h(\mp u_{3},-u_{2})}\Big)=0.
\end{align*}
Through the same argument as the two-magnon case, we showed that the spin chain length dependent term is exactly zero for the three-magnon case.

\section{On norms of structure constants \label{appg}}

In this appendix, we would like to give exact norm forms of the structure constants, especially for one nontrivial operator case. In ordinary, the correct structure constants including normalizations at tree-level are given by  
\begin{align*}
\left(\frac{C_{123}^{M\circ\circ}}{C_{123}^{\circ\circ\circ}}\right)^{2}=\frac{1}{{\cal N}^{(M)}}\left(\sum_{\ell_{12}+1\leq x_{1}<\cdots<x_{M}\leq L_{1}}\psi^{(M)}(x_{1},\cdots,x_{M})\right)^{2},
\end{align*}
where
\begin{align*}
{\cal N}^{(M)}=\sum_{1\leq x_{1}<\cdots\leq L_{1}}(\psi^{(M)})^{\mathfrak f}\psi^{(M)}.
\end{align*}
The subscript ${\mathfrak f}$ means the flipping operation introduced in \cite{Escobedo:2010xs}. For a few magnons, the normalization ${\cal N}^{(M)}$ can be exactly computed by the brute force method. On the other hand, the brackets part which is the main part of the structure constants has already given in section \ref{sec323}.

Let us begin by recalling the one-magnon wave function :
\begin{align*}
\psi^{(1)}(x)=e^{ip(x-\frac{1}{2})}+e^{2ipL}e^{-ip(x-\frac{1}{2})}.
\end{align*}
By flipping operation it can be easily found that ${\mathfrak f}:\psi^{(1)}\rightarrow e^{ipL}(\psi^{(1)})^{\ast}$.\footnote{For the closed chain, the operation is slightly different such as $(\psi_{\rm closed}^{(1)})^{\mathfrak f}=e^{ip(L+1)}\psi_{\rm closed}^{(1)}$.} 
Notice that we found a curious property of the open spin chain wave function that the conjugated wave function is the exactly same with the original wave function for the one-magnon : $(\psi^{(1)})^{\ast}=\psi^{(1)}$. 
By computing the summation of the square of the wave function, we can get the following insinuating identity for the norm ${\cal N}^{(1)}$:
\begin{align}
{\cal N}^{(1)}=\left({\cal M}(p)\right)^{2}(\partial_{u}\phi),
\end{align}
where the derivative is performed for the rapidity variable and $\phi$ is defined from the Bethe Yang equation for open spin-chain :\begin{align}
e^{i\phi}=e^{2ipL}.
\end{align}
Furthermore, remember that the factor ${\cal M}(p)$ came from the main part of the structure constant. 

For two-magnon case, the wave function is written by
\begin{align*}
\psi^{(2)}(x_{1},x_{2})=&g(x_{1},p_{1};x_{2},p_{2})+S(p_{2},p_{1})S(-p_{2},p_{1})e^{2ip_{1}L}g(x_{1},-p_{1};x_{2},p_{2})\notag\\
&+e^{2ip_{2}L}g(x_{1},p_{1};x_{2},-p_{2})+S(p_{2},p_{1})S(-p_{2},p_{1})e^{2i(p_{1}+p_{2})L}g(x_{1},-p_{1};x_{2},-p_{2}),
\end{align*}
where 
\begin{align*}
g(x_{1},p_{1};x_{2},p_{2})&\equiv{\cal A}(x_{1},p_{1}){\cal A}(x_{2},p_{2})+S(p_{2},p_{1}){\cal A}(x_{1},p_{2}){\cal A}(x_{2},p_{1}).
\end{align*}
The flipping operation can be written again by the original wave function such as
${\mathfrak f}:\psi^{(2)}\rightarrow e^{i(p_{1}+p_{2})L}S(p_{2},p_{1})(\psi^{(2)})^{\ast}$ and $(\psi^{(2)})^{\ast}=S(p_{1},p_{2})S(p_{1},-p_{2})e^{-2i(p_{1}+p_{2})L}\psi^{(2)}$. Thus, we can expect the general magnon case 
\begin{align*}
{\mathfrak f}:\psi^{(M)}\rightarrow (\psi^{(M)})^{\ast}\prod_{i<j}S(p_{j},p_{i})e^{i(p_{1}+\cdots+p_{M})L},
\end{align*}
and
\begin{align*}
(\psi^{(M)})^{\ast}=\prod_{i<j}S(p_{i},p_{j})S(p_{i},-p_{j})e^{-2i(p_{1}+\cdots+p_{2})L}\psi^{(M)}.
\end{align*}
From the summation for the wave function, we can obtain the norm for open spin-chain :
\begin{align}
\sum_{1\leq x_{1}<x_{2}\leq L_{1}}(\psi^{(2)})^{\ast}\psi^{(2)}=\left({\cal M}(p_{1}){\cal M}(p_{2})\right)^{2}{\rm det}(\partial_{u_{i}}\phi_{j}).
\end{align}
Here, the determinant ${\rm det}(\partial_{u_{i}}\phi_{j})$ is known as the Gaudin norm \cite{Bajnok:2015hla,Pozsgay}, where the $\phi_{j}$ is defined from the Bethe-Yang equation for the open spin-chain such as\footnote{We have $B(p)=1$ in our basis.}
\begin{align*}
e^{i\phi_{1}}=e^{2ip_{1}L_{1}}S(p_{2},p_{1})S(-p_{2},p_{1}),\\
e^{i\phi_{2}}=e^{2ip_{2}L_{1}}S(p_{1},p_{2})S(-p_{1},p_{2}).
\end{align*}
Therefore, the norm ${\cal N}^{(2)}$ can be given in terms of the Gaudin norm :
\begin{align}
{\cal N}^{(2)}=\left({\cal M}(p_{1}){\cal M}(p_{2})\right)^{2}{\rm det}(\partial_{u_{i}}\phi_{j})S(p_{2},p_{1})e^{i(p_{1}+p_{2})L}.
\end{align}

We finally expect that the norm for the multi-magnon is given as
\begin{align}
{\cal N}^{(M)}=\left(\prod_{i}{\cal M}(p_{i})\right)^{2}{\rm det}(\partial_{u_{i}}\phi_{j})\prod_{i<j}S(p_{j},p_{i})e^{i(p_{1}+\cdots+p_{M})L}. \label{normform}
\end{align}
We emphasize that we checked validity of (\ref{normform}) by numerically solving the Bethe ansatz equations. 

We finish with useful comments about numerical solutions of the BAEs. 
When we try to solve the BAEs numerically, solving such higher order polynomial equations takes quite long time even in symbolic programming. An efficient way is to take logarithm to original BAEs and to introduce the following technique:\footnote{For example, see \cite{pedrolecture} for more explanation.}
\begin{equation}
\frac{1}{i}\log\left(\frac{x+\frac{i}{2}}{x-\frac{i}{2}}\right) \rightarrow  -2\arctan{(2x)} \pm \pi \nonumber
\end{equation}
where the positive sign is taken for $x \geq 0$ and the negative sign is taken for $x < 0$. 
With this method, the Bethe equation can be rapidly solved in Mathematica.  

On the other hand, among solutions related to allowed mode numbers, there exist not alone physical solutions but also unphysical solutions.
Therefore, we have to take only physically admissible solutions when we would like to check (\ref{normform}) numerically.
Actually, the study on admissible solutions of various open BAEs has performed in \cite{rafael}.
In particular, the section 2 of \cite{rafael} treated the BAEs which are equivalent to our BAEs appearing in the Wilson loop with operator insertions.
Based on physical restrictions including reflection symmetry of the BAEs, the number of admissible Bethe rapidities is given as
\begin{equation}
{\cal N}(L, M) = \frac{L!}{M! (L-M)!}- \frac{L!}{(M-1)! (L-M+1)!},
\end{equation}
which only counts distinct, nonzero and not-equal solutions. In addition, the permutation of solutions is not counted in ${\cal N}(L, M)$.
Note that $L$ is the length of the open spin-chain and $M$ is the number of magnons.
Keeping this in mind, we could obtain some correct sets of Bethe rapidities. For these sets, we verified that (\ref{normform}) is indeed true.

\end{document}